\documentclass[12pt]{article}
\usepackage{amsmath,empheq}
\usepackage{amsfonts}
\usepackage[english]{babel}
\usepackage{amssymb}
\usepackage{amsthm}
\usepackage{graphicx}
\usepackage[a4paper]{geometry}
\usepackage{subfig}
\usepackage[colorlinks=true,linkcolor=blue,citecolor=blue]{hyperref}
\usepackage{amstext}
\usepackage{caption}
\usepackage{etoolbox}
\usepackage{makeidx}
\usepackage{amsfonts}
\usepackage{authblk} 
\usepackage{sectsty}
\usepackage{amscd}
\usepackage{mathrsfs}
\usepackage{dsfont}
\usepackage{enumitem} 
\usepackage[]{latexsym}
\usepackage{braket}
\usepackage{caption}

\usepackage{soul}
\usepackage{mathrsfs}
\usepackage{fancyhdr}
\usepackage{physics}
\usepackage{eufrak}
\usepackage[small,nohug,heads=littlevee]{diagrams}
\usepackage{circuitikz}
\diagramstyle[labelstyle=\scriptstyle]
\usepackage{xcolor}

	\newcommand{\ncd}{\newcommand}
	\newcommand{\vsp}{\vspace{0.4cm}}

	\ncd{\mrm}{\mathrm}
	\ncd{\beq} {\begin{equation}}
	\ncd{\eeq} {\end{equation}}

	\def\d{{\rm d}}
	
	\def\i{{\rm i}}

	\def\C{{\mathbb{C}}}
	\def\R{{\mathbb{R}}}
	
	\def\HP{{\mathbb{HP}}}

	\def\h{{\mathcal{H}}}

	\def\Ri{{\mathcal{C}}}

	\def\F{{\mathfrak{F}}}
	\def\X{{\mathfrak{X}}}
	\def\sl{{\mathfrak{sl}}}
	
	\def\su{{\mathfrak{su}}}

	\def\e{{\rm e}}
	
	\def\Tr{{\rm Tr}}

	\newcommand{\lag}{\mathfrak{L}}
	
\title{ Nonlinear Description of Quantum Dynamics.
\\ Generalized Coherent States. 
}

\author[1, 5]{H. Cruz-Prado}

\author[2, 3, 6]{G. Marmo}

\author[1, 7]{D. Schuch}

\author[4, 8]{O. Casta\~nos}

\providecommand{\keywords}[1]{\textbf{\textit{Keywords:}} #1}

\affil[1]{\textit{\footnotesize Institut f\"ur Theoretische Physik, Goethe-Universit\"at Frankfurt am Main, Max-von-Laue-Str. 1, D-60438 Frankfurt am Main, Germany.}}

\affil[2]{\textit{\footnotesize INFN-Sezione di Napoli, Complesso Universitario di Monte S. Angelo Edificio 6, via Cintia, 80126 Napoli, Italy.}}

\affil[3]{\textit{\footnotesize Dipartimento di Fisica ``E. Pancini'', Universit\`a di Napoli Federico II, Complesso Universitario di Monte S. Angelo Edificio 6, via Cintia, 80126 Napoli, Italy.}}

\affil[4]{\textit{\footnotesize Instituto de Ciencias Nucleares, Universidad Nacional Aut\'onoma de M\'exico, Apartado Postal  70-543, 04510 Ciudad de M\'exico, M\'exico.}}

\affil[5]{\footnotesize e-mail: \texttt{a062000@gmail.com}}

\affil[6]{\footnotesize e-mail: \texttt{marmo@na.infn.it}}

\affil[7]{\footnotesize e-mail: \texttt{schuch@em.uni-frankfurt.de}}

\affil[8]{\footnotesize e-mail: \texttt{ocasta@nucleares.unam.mx}}

\begin{document}

\maketitle

\abstract{
In this work it is shown that there is an inherent nonlinear evolution in the dynamics of the so-called generalized coherent states.
To show this, the immersion of a {\it classical manifold} into
the Hilbert space of quantum mechanics is employed. 
Then one may parametrize the time-dependence of the wave function through the variation of parameters in the classical manifold. 
Therefore, the immersion allows to consider the so-called \emph{principle of analogy}, i.e. using the procedures and structures available from the classical setting to employ them in the quantum framework.
}


%%%%%%%%%%%%%%%%%%%%%%%%%%%%%%%%%%%%%%%%%%%%%%%%%%%%%%%%%%%%%%%%%%%%%%%%%%%%%%%%%%%%%%%%%%%%%%%%%%%%%%%%%%%%%%%%

\section{Introduction}

One of the traditional problems in quantum mechanics is the study of Gaussian wave packets as analytic solution of the Schr\"odinger equation for Hamiltonians that are at most quadratic (or bilinear) in position and momentum variables, including problems whose Hamiltonian depends explicitly on time.

The rise of the Gaussian wave packets goes back to the seventies, when there was a large activity about the semiclassical limit of quantum mechanics; for example, many aspects of molecular quantum dynamics fall into the semiclassical domain~\cite{diracbook, feynmanbook}.
The study of such a domain goes from the expansion in $\hbar$, by introducing the description of the wave function in terms of phase and amplitude real functions~\cite{diracbook}, to the Feynman path integral approach~\cite{feynmanbook} or equivalently the linear time-dependent invariants, that is, time-dependent Gaussian wave packets~\cite{lewis1968, malkin1973, heller1975}. 
In addition, it has been found, for narrow wave packets and smooth potentials, that the wave packet feels only the first terms of the Taylor series expansion of the potential around its center~\cite{heller1975}.  
A review of a Hamiltonian model which has integrable solutions within the framework of the time-dependent Schr\"odinger equation is presented in~\cite{kleber1994}.  
The afore-mentioned works had impact on studies of several fields in the time-dependent domain; for example, in matter wave optics the changes in the energy spectrum of ultra-cold neutrons~\cite{hils1998}, in atom optics experiments using a time-modulated mirror~\cite{arndt1996} or studies of diffraction of a Bose-Einstein condensate in the time domain~\cite{colombe2005}.

As it has been already mentioned, there are several methods in the literature to study Gaussian wave packets. 
One of the most popular is the use of the so-called \emph{linear invariant operators} introduced by Lewis in Ref.~\cite{lewis1968} and Malkin, Man'ko and Trifonov in Ref.~\cite{malkin1973}. 
This approach is based in the construction of operators $\hat{I}$ such that their total derivative with respect to time is equal to zero and they are linear in the position and momentum variables, i.e. 
	\beq
	\hat{I} = a(t) \hat{q} + b(t) \hat{p}\, .
	\eeq 
The time-dependent functions $a(t)$ and $b(t)$ are determined from the invariance of the operator,  because this invariance defines a system of linear differential equation for these functions, which together with the initial conditions fix the functions $a(t)$ and $b(t)$; further details are given in~\cite{malkin1973}.  
One may demonstrate that these linear time-dependent invariants can be obtained from the Hamiltonian formulation of N\"other's theorem. 
The variations are related to the time-dependent functions $a(t)$ and $b(t)$~\cite{castanos1994}.

The solution of the Schr\"odinger equation is obtained from the fact that an invariant operator $\hat{I}$ is an operator that transforms every solution $|\psi \rangle$ of the Schr\"odinger equation into a solution $\hat{I} |\psi \rangle$ of the same equation, what is clear from the fact that $\hat{I}$ satisfies the equation
	\beq\label{Inv-def}
	\left[ \i \hbar \frac{\partial }{\partial t} - \hat{H}, \hat{I} \right] |\Psi, t \rangle = 0\, .
	\eeq
Therefore, with a single solution of the Schr\"odinger equation it is possible to find a family of them. 
In particular, it is possible to construct the invariant creation and annihilation operators $\hat{A}$ and $\hat{A}^\dagger$ with the usual commutation relation $[ \hat{A}, \hat{A}^\dagger ] = 1$. Thus we may define the eigenvalue equation for the annihilation operator
	\beq \label{Gen-Coh-St}
	\hat{A} | \alpha_{\tiny \mbox{Inv}} \rangle = \alpha_{\tiny \mbox{Inv}} | \alpha_{\tiny \mbox{Inv}} \rangle\, ,
	\eeq
where the eigenvalue $\alpha_{\tiny \mbox{Inv}} = \langle \alpha_{\tiny \mbox{Inv}} | \hat{A} | \alpha_{\tiny \mbox{Inv}} \rangle$ is a complex constant of motion and the eigenvector $ | \alpha_{\tiny \mbox{Inv}} \rangle$ is a solution of the Schr\"odinger equation.
To prove that the states $| \alpha_{\tiny \mbox{Inv}} \rangle$ are solution of the Schr\"odinger equation, let us consider the fiducial state $| 0 \rangle$ that is not only a solution of the Schr\"odinger equation but also satisfies $ \hat{A} | 0 \rangle = 0 | 0 \rangle$; thus, we may define the state 
	\beq \label{Gen-Coh-St-2}
	| \alpha_{\tiny \mbox{Inv}} \rangle = e^{ \alpha_{\tiny \mbox{Inv}} \hat{A}^\dagger 
	- \bar{\alpha}_{\tiny \mbox{Inv}} \hat{A}} | 0 \rangle \, ,
	\eeq 
where overbars denote the complex conjugate quantity.	
Then, because invariants transform solution of the Schr\"odinger equation into a solution of the same equation, as we mentioned before, from the expression \eqref{Gen-Coh-St-2} it should be clear that $| \alpha_{\tiny \mbox{Inv}} \rangle$ is a solution of the Schr\"odinger equation.
For example, for the one-dimensional system with Hamiltonian operator 
	\beq\label{2-Quad-Ham}
	\hat H 
	= \frac{1}{2} \,
	( \, \hat q \,\,\, , \,\,\, \hat p \, )
	\left(
	\begin{array}{ccc}
	H_1  &  V  \\
	& \\
	V &   H_2   \\    
	\end{array}
	\right)
	\left(
	\begin{array}{c}
	\hat q \\
	 \\
	\hat p
	\end{array}
	\right) \, ,
	\eeq
where $H_1$, $H_2$ and $V$ are possibly time-dependent functions, the bosonic linear invariant operators are
	\beq  \label{Inv-bosonic-op}
	\hat{A} = \frac{\i}{\sqrt{2\hbar}} \left( P \, \hat{q} - Q \, \hat{p} \right)  \,
	\quad
	\text{and}
	\quad
	\hat{A}^\dagger =  \frac{- \i}{\sqrt{2\hbar}} \left( \bar{P} \, \hat{q} -  \bar{Q} \, \hat{p} \right)\, .
	\eeq
Now, in order to fulfil $[\hat{A} , \hat{A}^\dagger] = 1$ the time-dependent complex functions $Q$ and $P$ are constrained to satisfy 
	\beq \label{QP-Constrain}
	\bar{Q} P - Q \bar{P} = 2 \, \i \, .
	\eeq 
On the other hand, from the invariance of the operators one directly obtains that these functions obey the linear system of differential equations
 	\beq
	\label{eqno8}
	\left(
		\begin{array}{ccc}
			\dot{Q} \\  
			\\
			\dot{P}
		\end{array}
	\right)
	 = 
	\left(
		\begin{array}{cc}
		V   & H_2 \\		
		& \\
		- H_1 & - V \\
		\end{array}
	\right)
	\left(
		\begin{array}{ccc}
			Q \\  
			\\
			P
		\end{array}
	\right) \, .
	\eeq
Then, the Gaussian solution of the Schr\"odinger equation may be obtained with the help of the linear invariant operators and establishing the definition \eqref{Gen-Coh-St-2} in the position representation, 
	\beq \label{Gaussian-WP}
	\langle q \, | \, \alpha_{\tiny \mbox{Inv}} \rangle = \frac{1}{ (\pi \, \hbar)^{1/4}} \frac{1}{\sqrt{Q}} 
	\exp
	\left\{
	\frac{\i}{2 \hbar} \, \frac{P}{Q} \, ( q - \langle \hat{q} \rangle)^2 
	+ \frac{\i}{\hbar} \langle \hat{p} \rangle ( q - \langle \hat{q} \rangle) 
	+ \frac{\i}{2 \hbar} \,  \langle \hat{q} \rangle \,  \langle \hat{p} \rangle 
	\right\} \, .
	\eeq
The state $| \alpha_{\tiny \mbox{Inv}} \rangle$ defined by the condition \eqref{Gen-Coh-St} is called in the literature \emph{generalized coherent state}~\cite{dodonov2003}, for which, in contrast to the standard coherent state, the correlation between $\hat{p}$ and $\hat{q}$ may be different from zero, and hence these states are such that they minimize the so-called \emph{Robertson--Schr\"odinger uncertainty relation}, i.e.,
	\beq \label{Sch-Rob-Id}
	\sigma_q \sigma_p - \sigma_{qp}^2 = \frac{\hbar^2}{4} \, ,
	\eeq 
where 
	\beq \label{sec-mom-def}
	\sigma_{q} = \langle \hat{q}^2 \rangle - \langle \hat q \rangle^2\, , 
	\quad 
	\sigma_{p} = \langle \hat{p}^2 \rangle - \langle \hat{p} \rangle^2
	\quad
	\text{and}
	\quad
	\sigma_{ q p} = \frac{1}{2}\langle \hat{q}  \hat{p} + \hat{p}  \hat{q} \rangle 
	- \langle \hat{q} \rangle \langle \hat{p} \rangle \, .
	 \eeq

The linear invariant approach for the study of Gaussian wave packets depends on the solution of the linear equations of motion defined for the time-dependent functions $a(t)$ and $b(t)$, which may be complex or real functions. 
However, there are different approaches that involve non-linear differential equations, specifically an evolution described by  the solutions of the Ermakov--Lewis and Riccati equations~\cite{ lewis1968, carinena1998, cruz_schuch_castanos_rosas-ortiz-1, cruz_schuch_castanos_rosas-ortiz-2, schuch2018}.
For example, according to \emph{the Wei--Norman method} \cite{wei1963} employed in \cite{carinena1998} there are complex functions $\Ri_k(t)$, with $k=1,2,3$ and $\Ri_k(0) = 0$, such that the unitary evolution operator may be expressed as
	\begin{align}\label{Ev-op-1}
	\hat{U}(t) & = \e^{\frac{\i}{2 \hbar} \, \Ri_1(t)\hat{q}^2}
	\e^{\frac{\i}{2 \hbar} \, \Ri_2(t) [\hat{q}, \hat{p} ]_{+}} \
	\e^{\frac{\i}{2 \hbar} \, \Ri_3(t) \hat{p}^2} \nonumber \\
	& = \hat{U}_{\Ri_1}(t) \hat{U}_{\Ri_2}(t) \hat{U}_{\Ri_3}(t) \, .
	\end{align}
The time-dependent functions $\Ri_k$ are determined by direct substitution of the evolution operator into the equation of motion
	\beq\label{Eq-Mot-TDO}
	\i \hbar \frac{\d \, \hat{U}}{\d t} = \hat{H}  \hat{U} \, 
	\eeq
with initial condition $\hat{U}(0) = \hat{1}$. 
For the Hamiltonian given in \eqref{2-Quad-Ham} one may prove that the complex functions $\Ri_k$ obey the system of differential equations
	\begin{align}
	\dot{\Ri}_1 & = - H_2 \, \Ri_1^2 - 2 V \, \Ri_1 - \, H_1\, , \label{Ric-Un} \\
	\dot{\Ri}_2 & = - H_2 \, \Ri_1 - V\, , \label{com1-eq1} \\
	\dot{\Ri}_3 & =  - \e^{ 2 \Ri_2} \, H_2\, . \label{com2-eq1}
	\end{align}
Notice that the evolution operator \eqref{Ev-op-1} can be constructed with the operators $\hat{U}_{\Ri_k}(t)$ in different order, giving rise to different Riccati differential equations; however, all of them are connected by M\"obius transformations, for details see Ref.~\cite{carinena1998}.
Furthermore, according to Ref.~\cite{schuch2018}, the Gaussian wave function solution of the Schr\"odinger equation may be expressed as
	\begin{eqnarray}
	\label{G-WP-Ri-q}
	\psi(q) = \frac{1}{ (\pi \, \hbar)^{1/4}}
	\exp\left\{
	\frac{\i}{2 \hbar} \, \Ri \, ( q - \langle \hat{q} \rangle)^2 
	+ \frac{\i}{\hbar} \langle \hat{p} \rangle ( q - \langle \hat{q} \rangle) 
	+ \frac{\i}{2 \hbar} \langle \hat{q} \rangle \,  \langle \hat{p} \rangle
	-\frac{1}{2} \int^t [H_2 \, \Ri(t') + V ]\d t'
	\right\} \, .
	\end{eqnarray}
where $\Ri$ is a complex time-dependent function with imaginary part strictly positive and different from zero. 
In addition, this function obeys the nonlinear Riccati equation
	\beq \label{Osc-Ric-Q}
	\dot{\Ri} + H_2 \, \Ri^2  + 2 V \, \Ri + H_1 = 0 \, ,
	\eeq
which is the same Riccati equation as given in Eq.~\eqref{Ric-Un} for the  determination of the unitary evolution operator.

The connection between linear and nonlinear evolution descriptions of the generalized coherent states has been studied already~\cite{cruz_schuch_castanos_rosas-ortiz-1, schuch2018}  where the transformations that connect the different descriptions of the time-dependent solutions of quadratic Hamiltonian are explicitely established.  

In a framework different from the before mentioned approaches for the study of the generalized coherent states, here this problem will be analized from a geometrical point of view.
This means that the geometrical space where the different evolutions take place will be established  and it is shown how they are \emph{immersed} in the Hilbert space $\h$, as well as how the connection among the different descriptions is established.
Moreover, it is demonstrated that each space has a symplectic structure such that the dynamics on these spaces are actually Hamiltonian.
Then the previously mentioned results are not only deduced from a geometrical perspective but it is also shown that there are several nonlinear descriptions involved in the dynamics of the generalized coherent states.

The structure of this paper is as follows. 
In order to make the paper self-containedn in Section \ref{Section-2} we introduce some geometrical aspects involved in the usual description of the generalized coherent states together with its corresponding standard linear evolution. 
In Section \ref{Section-3}, it is demonstrated that from the before mention linear evolution it is possible to obtain another Hamiltonian, but nonlinear evolution, defined in a Hyperboloid manifold, described completely by hyperbolic coordinates $(\tau, \varphi)$, which in Quantum Optics are known as \emph{squeezing parameters}.
Section \ref{Section-4} is devoted to the Hamiltonian nonlinear Riccati evolution, where it is proven that the evolution can take place in two types of spaces, the \emph{Poincar\'e Disk} and in the \emph{Siegel upper half plane}.
To apply the formalism developed in the previous sections, in Section \ref{Section-5} \emph{the degenerate parametric amplifier}, which is an optical device with one effective electromagnetic mode, is analized.
Finally, in Section \ref{Section-6} a summary of the results is given and future directions of our investigation are highlighted.

%%%%%%%%%%%%%%%%%%%%%%%%%%%%%%%%%%%%%%%%%%%%%%%%%%%%%%%%%%%%%%%%%%%%%%%%%%%%%%%%%%%%%%%%%%%%%%%%%%%%%%%%%%%%%%%%%%%%%%%%%%%%%%%%%%%%%%%%%%%%%%%%%%%%%%%%%%%%%%%%%%%%%%%%

\section{ Linear Description of Generalized Coherent States}

\label{Section-2}

\begin{figure}[! h]
	\begin{center}
	\includegraphics[width = 15 cm]{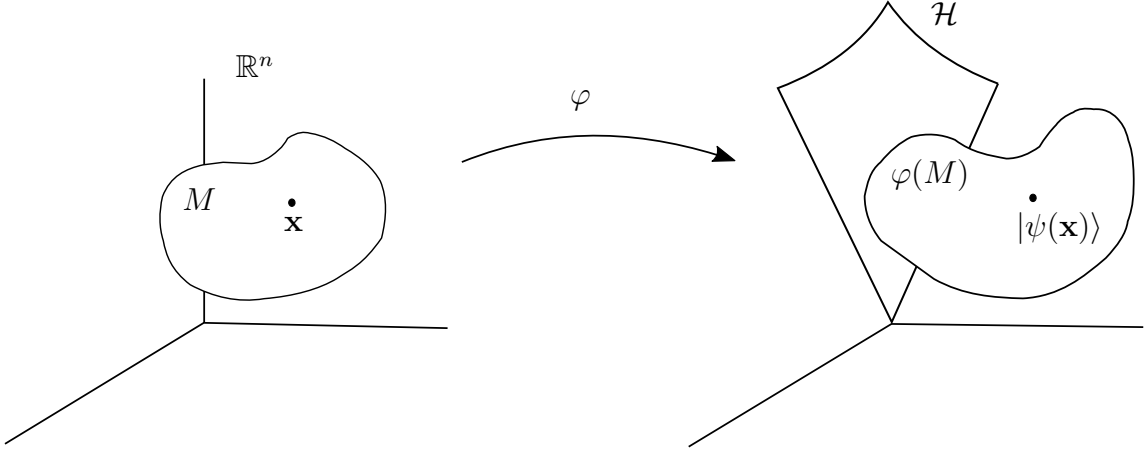}
	\put(-80,160){$\h$}
	\put(-95,100){$\varphi(M)$}
	\put(-322,82){$\mathbf{x}$}
	\put(-215,130){$\varphi$}
	\put(-340, 140){$\R^n$}
	\put(-360, 90){$M$}
	\put(-48,80){$ | \psi(\mathbf{x}) \rangle $}
\caption{Representative picture of the immersion $\varphi$ of the submanifold $M \in \R^n$ into the Hilbert space $\h$, leading to the submanifold $\varphi(M) \subset \h$.}
\label{Fig-1}
\end{center}
	\end{figure}

To start the study some geometrical aspects of the usual description of the generalized coherent states are established, for a more complete review see~\cite{kramer1981, esposito2004, carinena2015geometry}. 
So, in quantum mechanics it is possible to immerse a manifold $M \subset \R^n$ in the Hilbert space $\h$ of a physical system injectively, i.e.,
	\beq
	\varphi : M \to \h : \mathbf{x} \mapsto | \psi(\mathbf{x}) \rangle \, ,
	\eeq
such that $\varphi(M)$ is a submanifold of $\h$~\cite{ciaglia2017}, see Fig.~\ref{Fig-1}. 
Recall that an \emph{immersion} $\varphi$ is a differentiable map in which $T\varphi$ is injective, i.e. 
$T_\mathbf{x} \psi : TM_\mathbf{x} \to T\h_{| \psi \rangle}$ is an injective function at every point $\mathbf{x} \in M$.
Hence, one may parametrize the time dependence of the wave function $| \psi \rangle$ through the variation of the parameters $\mathbf{x}$, whose physical significance will depend on the problem at hand.  

Let us now show for the case of the generalized coherent states that we may define an immersion.
For simplicity, here the case of the generalized coherent states for a one-dimensional quantum system is considered, the generalization to more dimensions is straightforward; then, the immersion is constructed as 
follows~\cite{kramer1981, ciaglia2017}.

Given a fiducial normalized state $| 0 \rangle$ in the Hilbert space $\h$, one employs \emph{the strongly continuous map} $\hat{D}$ to immerse the complex plane $\C$ in the Hilbert space, 
	\beq \label{immersion}
	i : \alpha \in \C \mapsto | \alpha \rangle \in \h
	\quad
	\text{with}
	\quad
	| \alpha \rangle := \hat{D}(\alpha) \, | 0 \rangle.
	\eeq 
The operator $\hat{D}(\alpha)$ is the well-known \emph{displacement operator} given by
	\beq
	\hat{D}(\alpha) = \e^{\alpha \, \hat{a}^\dagger - \bar{\alpha} \, \hat{a}} \, ,
	\eeq 
where $\hat{a}$ is the annihilation and $\hat{a}^\dagger$ the creation boson operators with $\alpha$ and $\bar{\alpha}$ their respective expectation values.
Now, considering the Hilbert space of quadratically integrable functions $\lag^2(\R, \d q)$, the immersed submanifold $i(\C)$ corresponds to the \emph{Gaussian wave packets}, i.e. one has the well-known immersion~\cite{ciaglia2017, ciaglia2019}
	\beq \label{inm-qua-fun}
	i : \C \to \lag^2(\R, \d q) : \alpha \mapsto \psi( \alpha, q ) \, .
	\eeq
It should be clear that this immersion corresponds to a pure state with a fixed dependence on the parameter $\alpha$. 
It is clear that the superposition of generalized coherent states does not belong to $i(\C)$, because such a superposition is not a Gaussian wave packet.
An important consequence of this is that the subset $i(\C)$ of the Hilbert space is a nonlinear space. 
As a final remark, it is not difficult to see from the definition of the displacement operator that 
	\beq \label{Weyl-cond}
	 \hat{D}^\dagger(\alpha) \, \hat{D}^\dagger(\beta) \, \hat{D}(\alpha) \, \hat{D}(\beta)
	 =   \, \e^{\frac{\i}{\hbar} \, \omega(\alpha, \beta) } \, ,
	\eeq	
where $\omega(\alpha, \beta)$ is the standard symplectic form $\omega = \d x \wedge \d y$ in $\C$, which is represented by the matrix
	\beq \label{complex-str}
	J =  \left(
	\begin{array}{ccc}
	0 & 1 \\
	-1 & 0 \\   
	\end{array}
	\right) \, .
	\eeq 
Thus, being $\alpha = x + \i \, y$ and $\beta = x' - \i \, y'$ we have that
	\beq
	\omega( \alpha , \beta ) = ( x \, , \, y )
	\left(
	\begin{array}{cc}
	0  &  1   \\
	 -1 &  0   \\   
	\end{array}
	\right)
	\left(
	\begin{array}{c}
	 x'   \\
	 y'
	\end{array}
	\right) 
	= x \ y' - x' \, y \, .
	\eeq
The relation \eqref{Weyl-cond} is known as the \emph{Weyl form} of the commutation relations and together with the immersion \eqref{immersion} allows to see that we are actually dealing with a \emph{Weyl system}~\cite{weyl1950theory, ercolessi2010equations}, i.e., a projective unitary representation of the Abelian vector group $\C$ in the Hilbert space.
An important consequence of dealing with a Weyl system is the fact that we are avoiding domain problems of the Hermitian operators, that is, the domain of the creation and annihilation operator are well defined in the submanifold $i(\C) \subset \h$.

\vsp

Now that the kinematics involved in the generalized coherent states has been established, we are interested in the dynamical properties of the system.
So, the evolution of a pure state $| \alpha \rangle \in \h$ is given by the Schr\"odinger equation 
	\beq \label{Sch-Eq}
	\i \, \hbar \frac{\partial \, | \alpha \rangle }{\partial \, t} = \hat{H} \, | \alpha \rangle
	\eeq
with $\hat{H}$ being the Hamiltonian operator. 
Thus assuming that the solution of the system can be extended to the whole time, let be $\hat{U}_t : \h \to \h$ the one-parameter group of unitary transformations associated with the Schr\"odinger equation \eqref{Sch-Eq}. 
In this work only Hamiltonian operators quadratic in the position and momentum variables are considered. 
For these cases, it has been shown in \cite{mehta1966, dodonov1975} that if the initial state is a generalized coherent state, the final state is also a generalized coherent state, 
	\beq
	\hat{U}_t | \alpha_0 \rangle = | \alpha(t) \rangle \, , 
	\eeq
what implies that the submanifold $i(\C) \subset \h$ is invariant with respect to the flow $U_t$.

On the other hand, it is well known that the field of complex numbers $\C$ may be identified with the Euclidian space $\R^2$, i.e. if $\alpha = x + \i \, y$; $x,y \in \R$, then the corresponding vector in $\R^2$ is represented by $(x,y)$ and then it is immediate to check that the group property is satisfied.
This identification allows to define a symplectic structure on $\C$, i.e., $\C$ is naturally endowed with a symplectic form $\omega$; then, one may associate to every differentiable real function $H \in \F(\C)$ the vector field $X_H$, which is described by the interior product $i_{X_H}$ acting on the symplectic form $\omega$, that is,
	\beq
	i_{X_H} \omega =\omega(X_H,\cdot)= - \d H\, .
	\eeq
\emph{The Hamiltonian vector field}  $X_H$ defines the dynamics of the system, and its integral curves are fixed by the solutions of the Hamiltonian equations of motion.
In addition, one may introduce the Poisson brackets, namely given the real functions $F, G \in \F(\C)$ with associated vector fields $X_F$ and $X_G$ the Poisson bracket is defined as
	\beq \label{Def-Poisson}
	\{ F , G \} = \omega (X_F, X_G) \, .
	\eeq

In general, the solution of the Hamiltonian equations of motion may be expressed as $\alpha(t) = \Phi_t \alpha_0$,
where $\Phi_t: \C \to \C$ is one-parameter group of symplectic transformations, i.e. $\Phi_t \in \mbox{Sp}( 2n, M)$\footnote{
The $\mbox{Sp}(2n,M)$ is the symplectic group of degree $2n$ over $M$ defined as
	\begin{equation*}
	\mbox{Sp}(2n, M) = \{ S \in \mathbb{M}_{2n \times 2n} \, | \, S^{\tiny \mbox{T}} \, J \, S = J \} \, ,
	\end{equation*}
where $J$ is the symplectic matrix
	\begin{equation*} \label{complex-str}
	J =  \left(
	\begin{array}{ccc}
	0 & \mathbb{I}_{n \times n} \\
	- \mathbb{I}_{n \times n} & 0 \\   
	\end{array}
	\right) \, .
	\end{equation*}
}. 
Therefore, the connection between the unitary evolution $\hat{U}_t$ and the (canonical) symplectic evolution $\Phi_t$ is obtained by means of the immersion \eqref{immersion}, namely
	\beq
	i \circ \Phi_t  = i \circ \hat{U}_t \, . 
	\eeq

To avoid future confusions, let us note that in general for the description of a dynamical systems, classical or quantum, one may use complex valued vector fields and functions, but the carrier space remains as a real differential manifold.
This means that the exterior differential calculus is done on real manifolds, but applied to complex valued objects. This is the perspective that is consider along this work.

\vsp

\subsubsection*{Quantum parametric oscillator}

As an example, let us consider the Schr\"odinger equation of a Hamiltonian operator quadratic in the position and the momentum variables given in Eq.~\eqref{2-Quad-Ham}. 
This Hamiltonian can be equivalently expressed in terms of bosonic operators
	\beq  \label{bosonic-op}
	\hat{a} = \frac{\i}{\sqrt{2\hbar}} \left( P \, \hat{q} - Q \, \hat{p} \right)  \,
	\quad
	\text{and}
	\quad
	\hat{a}^\dagger =  \frac{- \i}{\sqrt{2\hbar}} \left( \bar{P} \, \hat{q} -  \bar{Q} \, \hat{p} \right)\, ,
	\eeq
where in order to have $[ \hat{a}, \hat{a}^\dagger ] = 1$,  $(Q , P)$ are points in the manifold $M$ defined as
	\beq \label{M-def}
	M = \{ (Q, P) \in \C^2 \, | \, \bar{Q} P - Q \bar{P} = 2 \, \i  \} \, .
	\eeq	
Thus in terms of the creation and annihilation operators we have the equivalent Hamiltonian
	\beq\label{2-Al-Ham}
	\hat H 
	= \frac{\hbar}{4} \,
	( \, \hat{a} \,\,\, , \,\,\, \hat{a}^\dagger \, )
	\left(
	\begin{array}{ccc}
	\bar{G}  &  W  \\
	      & \\
	W  &  G   \\    
	\end{array}
	\right)
	\left(
	\begin{array}{c}
	\hat{a} \\
	 \\
	\hat{a}^\dagger
	\end{array}
	\right)	\, ,
	\eeq
where the time-dependent functions $G$ and $W$ are connected with $H_1$, $H_2$ and $V$ by
	\begin{align}
	G & = H_1 Q^2 + 2 V Q P + H_2 P^2 \, , \\
	W & = H_1 |Q|^2 + V(Q \bar{P} + P \bar{Q} )+ H_2 |P|^2 \, .
	\end{align}

Let us now explicitly establish the immersion \eqref{inm-qua-fun}. So, the normalized state $| 0 \rangle $ is defined by the usual condition: $\hat{a} | 0 \rangle = 0 | 0 \rangle$.
Thus, the wave function of the ground state is obtained by means of the relation $\langle q | \hat{a} | 0 \rangle = 0$,
which in the position representation defines a partial differential equation for the ground state wave function, whose normalized solution is given by
	\beq \label{Ground-WP}
	\psi(0 , q) = \frac{1}{ (\pi \, \hbar)^{1/4}} \frac{1}{\sqrt{Q}}
	\, \exp\left\{
	\frac{\i}{2 \hbar} \, \frac{P}{Q} \, q^2 
	\right\} \, .
	\eeq
Now, the immersion defined in \eqref{inm-qua-fun} allows to establish a correspondence between the complex number
	\beq \label{alpha-coordinates}
	\alpha = \langle \alpha| \hat{a} | \alpha \rangle 
	= \frac{\i}{\sqrt{2\hbar}} \left( P \, \langle \hat{q} \rangle 
	- \, Q \, \langle \hat{p} \rangle \right)
	\eeq
and the Gaussian wave packet
	\begin{align} \label{Gaussian-WP}
	\psi(\alpha, q) & = \langle q | \hat{D}(\alpha) | 0 \rangle  \nonumber \\
	& = \frac{1}{ (\pi \, \hbar)^{1/4}} \frac{1}{\sqrt{Q}} 
	\exp
	\left\{
	\frac{\i}{2 \hbar} \, \frac{P}{Q} \, ( q - \langle \hat{q} \rangle)^2 
	+ \frac{\i}{\hbar} \langle \hat{p} \rangle ( q - \langle \hat{q} \rangle) 
	+ \frac{\i}{2 \hbar} \,  \langle \hat{q} \rangle \,  \langle \hat{p} \rangle 
	\right\} \, 
	\end{align}
in the Hilbert space $\lag^2(\R, \d q)$. 
Furthermore, it should be clear that one may also consider the immersion 
	\beq 
	i : \C \to \lag^2(\R, \d p) : \alpha \mapsto \tilde{\psi}(\alpha, p ) \, .
	\eeq 
where in this case the Gaussian wave packet in the momentum representation is given by
	\begin{align} \label{P-Gaussian-WP}
	\tilde{\psi}(\alpha, p) & = \langle p | \hat{D}(\alpha) | 0 \rangle  \nonumber \\
	& = \frac{1}{ (\pi \, \hbar)^{1/4}} \sqrt{\frac{\i}{P}}
	\exp
	\left\{
	- \frac{\i}{2 \hbar} \, \frac{Q}{P} \, ( p - \langle \hat{p} \rangle)^2 
	- \frac{\i}{\hbar} \langle \hat{q} \rangle ( p - \langle \hat{p} \rangle) 
	- \frac{\i}{2 \hbar} \,  \langle \hat{q} \rangle \,  \langle \hat{p} \rangle 
	\right\} \, .
	\end{align}

With the help of the Gaussian wave function $\psi(\alpha, q)$ (or its Fourier transform $\tilde{\psi}(\alpha, p)$), one may prove that there is a direct connection between the uncertainties $\sigma_{q}$, $\sigma_{p}$ and their correlation $\sigma_{ q p}$, all of them defined in \eqref{sec-mom-def}, with the complex quantities $(Q , P)$ by
	\beq \label{Unc-QP}
	\sigma_q = \frac{\hbar}{2} |Q|^2 \, ,
	\quad
	\sigma_p =  \frac{\hbar}{2} |P|^2 
	\quad
	\text{and}
	\quad
	\sigma_{qp} = \frac{\hbar}{4} (P \, \bar{Q} + Q \, \bar{P})  \, .
	\eeq
Establishing the covariance matrix $\Sigma $, one has that it may be factorized as
	\beq \label{Sigma-Com}
	\Sigma = \frac{2}{\hbar}
	\left(
	\begin{array}{ccc}
	\sigma_{qp} &  \sigma_{p}  \\
	& \\
	-\sigma_{q} & - \sigma_{qp}   \\   
	\end{array}
	\right)
	=
	\frac{1}{2}
	\left(
	\begin{array}{ccc}
	\bar{P} & P  \\
	& \\
	-\bar{Q} & - Q   \\   
	\end{array}
	\right)
	\left(
	\begin{array}{ccc}
	Q & P  \\
	& \\ 
	\bar{Q} & \bar{P}   \\   
	\end{array}
	\right)
	 \, .
	\eeq	
Therefore, the quantities $(\sigma_{q}, \sigma_{p}, \sigma_{qp})$, along with their constraint \eqref{Sch-Rob-Id} determine the values of $(Q, P)$. 
Note that the minimization of the Robertson--Schr\"odinger uncertainty relation directly introduces the constraint in Eq.~\eqref{QP-Constrain} for $(Q, P)$.

Let us now address the dynamical properties of this example.
Considering the unitary evolution in the Hilbert space $\lag^2(\R, \d q)$ generated by the Hamiltonian operator \eqref{2-Quad-Ham} (or the Hamiltonian \eqref{2-Al-Ham}), one has then that the associated symplectic evolution in $\C \approx \R^2$ is determined by the Hamiltonian dynamics $X \in \X(T^\ast \C)$ defined by 
	\beq \label{alph-ev}
	i_{X} \omega = - \, \d e_{\hat H} \, ,
	\eeq
where $\omega$ is the symplectic form, which in terms of the complex coordinates given in Eq.~\eqref{alpha-coordinates}, has the form
	\beq \label{sym-form}
	\omega = \d \langle \hat{p} \rangle \wedge \d \langle \hat{q} \rangle 
	=  \i \, \hbar \, \d \bar{\alpha}  \wedge \d \alpha  \, ,
	\eeq
and the Hamiltonian function $e_{\hat H}$ corresponds to the expectation value of the Hamiltonian, i.e., 
	\beq \label{exp-val}
	e_{\hat H} = \langle \alpha | \hat{H} | \alpha \rangle
	= \frac{\hbar}{4} \,
	( \, \alpha \,\,\, , \,\,\, \bar{\alpha} \, )
	\left(
	\begin{array}{ccc}
	\bar{G}  &  W  \\
	      & \\
	W  &  G   \\    
	\end{array}
	\right)
	\left(
	\begin{array}{c}
	\alpha \\
	 \\
	\bar{\alpha}
	\end{array}
	\right)	\, .
	\eeq	
So, inserting the expressions of the symplectic form and the expectation value in coordinates into the definition \eqref{alph-ev}, one may prove that 
	\beq
	X = \frac{\i}{2} \, \left( \bar{G} \, \alpha + W \, \bar{\alpha} \right) \frac{\partial}{\partial \bar{\alpha} }
	-  \frac{\i}{2} \, \left( G \, \bar{\alpha} + W \, \alpha \right) \frac{\partial}{\partial \alpha} \, ,
	\eeq
whose integral curves are given by the solutions of the Hamiltonian equations of motion
	\begin{align} \label{alpha-Ham}
	\dot{\alpha} & = - \frac{\i}{\hbar} \, \frac{\partial \, e_{\hat H} }{\partial \bar \alpha} 
	= - \frac{\i}{2} \, \left( G \, \bar{\alpha} + W \, \alpha \right) \, , \nonumber \\
	\dot{\bar{\alpha}} & = \frac{\i}{\hbar} \, \frac{\partial \, e_{\hat H} }{\partial \alpha}  
	= \frac{\i}{2} \, \left( \bar{G} \, \alpha + W \, \bar{\alpha} \right) \, .
	\end{align}
Moreover, one may use the Poisson brackets deflined by the relation \eqref{Def-Poisson}, given the expectation values $e_A$ and $e_B$ associated with the observables $\hat{A}$ and $\hat{B}$. The corresponding Poisson bracket is then given by
	\begin{align}
	\left\{ e_A , e_B \right\} & = 
	\frac{\i}{\hbar} \, \left[  \frac{\partial e_A}{\partial \alpha} \frac{\partial e_B}{\partial \bar{\alpha} }
	-   \frac{\partial e_A}{\partial \bar{\alpha}} \frac{\partial e_B}{\partial \alpha}
	\right] \nonumber \\ 
	& =  \frac{\partial e_A}{\partial \langle p \rangle} \frac{\partial e_B}{\partial \langle q \rangle }
	-   \frac{\partial e_A}{\partial \langle q \rangle} \frac{\partial e_B}{\partial \langle p \rangle}
	\, .
	\end{align}
This definition allows to introduce the evolution of the expectation value $e_A$ of an arbitrary observable $\hat{A}$ given by
	\beq
	\frac{\d \, e_A}{\d t} = \left\{ e_H , e_A \right\} \, .
	\eeq

Note that the formalism presented may be extended to the time-dependent case, i.e., for the evolution of functions which explicitly depend on time. 
For such cases, as usual, one simply considers the extended space $T^\ast \C \times \R$ as a carrier space, such that the evolution of the time-dependent expectation value $e_A$ is described by means of time-dependent vector fields
	\beq
	\frac{\d \, e_A}{\d t} = \left\{ e_H , e_A \right\} + \frac{\partial e_A}{\partial t} \,.
	\eeq

\vsp

Let us observe that until here there is nothing new, because the Hamiltonian equations \eqref{alpha-Ham} are simply the well-known \emph{Ehrenfest theorem}, which establishes that the maximum of the Gaussian wave packet follows the classical trajectories of motion 
	\begin{align} \label{Ehrenfest-Teo}
	\dot{\langle q \rangle} & = \frac{\partial \, e_{\hat H} }{\partial \langle p \rangle} 
	= V \langle q \rangle + H_2 \langle p \rangle \, , \nonumber \\
	\dot{\langle p \rangle} & = -  \frac{\partial \, e_{\hat H} }{\partial \langle q \rangle}  
	= - H_1 \langle q \rangle - V \langle p \rangle \, .
	\end{align}
In addition, by construction the complex quantities $(Q, P)$ are constant parameters; thus, from the relations in Eq. \eqref{Unc-QP}, it is clear that we have considered Gaussian wave packets which preserve the values of the second moments $(\sigma_{q}, \sigma_{p}, \sigma_{qp})$ during their evolution. 

%\textcolor{red}{ Recordar el significado de $H_1$, $H_2$, y $V$ (ver expresión (5))}

In order to consider the general situation, i.e., to have Gaussian wave packets with time-dependent second moments, it is necessary to introduce the dynamics in the manifold $M$, defined in Eq.~\eqref{M-def}.
To define such a dynamics we use the vector field $X_M$, given by a vector field in the manifold $M$, i.e. $X_M \in \X(M)$.  As the manifold $M$ has a symplectic structure of the form 
	\begin{align} \label{Sym-Comp}
	\omega_M & = \d \bar{P} \wedge \d Q + \d P \wedge \d \bar{Q} \nonumber \\
	& = 2 \, \d P_{\tiny \mbox{R}} \wedge \d Q_{\tiny \mbox{R}} 
	+ 2 \, \d P_{\tiny \mbox{I}} \wedge \d Q_{\tiny \mbox{I}} \, ,
	\end{align}
considering $Q = Q_{\tiny \mbox{R}} + \i \, Q_{\tiny \mbox{I}}$ and $P = P_{\tiny \mbox{R}} + \i \, P_{\tiny \mbox{I}}$. 
Then the dynamics is  symplectic, $i_Y \omega_M = - \d H_M$, where the Hamiltonian function corresponds to
	\beq
	H_M 
	=
	( \, \bar Q \,\,\, , \,\,\, \bar P \, )
	\left(
	\begin{array}{ccc}
	H_1  &  V  \\
	& \\
	V &   H_2   \\    
	\end{array}
	\right)
	\left(
	\begin{array}{c}
	Q   \\
	 \\
	P
	\end{array}
	\right) \, .
	\eeq
Therefore the vector field is given by
	\beq \label{Dynamics-M}
	X_M = (V Q + H_2 P) \frac{\partial}{\partial Q} - ( H_1 Q + V P) \frac{\partial}{\partial P}
	+  (V \bar Q + H_2 \bar P) \frac{\partial}{\partial \bar Q} 
	- ( H_1 \bar Q + V \bar P) \frac{\partial}{\partial \bar P} \, ,
	\eeq
with linear Hamiltonian equations of motion
	\beq \label{Eq-Com-Var}
	\left(
		\begin{array}{ccc}
			\dot{Q} \\  
			\\
			\dot{P}
		\end{array}
	\right)
	 = 
	\left(
		\begin{array}{cc}
		V   & H_2 \\		
		& \\
		- H_1 & - V \\
		\end{array}
	\right)
	\left(
		\begin{array}{ccc}
			Q \\  
			\\
			P
		\end{array}
	\right) \, .
	\eeq
Note that these are the equations of motion obtained in the linear invariant approach, see Eq.~(\ref{eqno8}). 
This is not a mere coincidence as it is seen next.
Taking into account the evolution of the complex quantities $(Q, P)$ in the definition \eqref{alpha-coordinates}, one obtains the time-dependent function
	\beq \label{Inv-St}
	\alpha_{\tiny \mbox{Inv}}(t) = \frac{\i}{\sqrt{2\hbar}} \left( P(t) \, \langle \hat{q} \rangle 
	- \, Q(t) \, \langle \hat{p} \rangle \right)\, ,
	\eeq
which obeys 
	\beq
	\frac{\d \alpha_{\tiny \mbox{Inv}}}{\d t} = \left\{ e_H ,  \alpha_{\tiny \mbox{Inv}} \right\} 
	+ \frac{\partial  \alpha_{\tiny \mbox{Inv}}}{\partial t} = 0 \, ,
	\eeq
i.e., $\alpha_{\tiny \mbox{Inv}}(t)$ is a constant of motion of the dynamics defined by the Hamiltonian function $e_H$ in Eq~\eqref{exp-val}. 

The bosonic creation $\hat{A}^\dagger$ and annihilation $\hat{A}$ operators for the time-dependent case are given by
\begin{equation}
\hat{A} = \frac{\i}{\sqrt{2\hbar}} \left( P(t) \, \hat{q} - Q(t) \, \hat{p} \right)  \,
	\quad
	\text{and}
	\quad
	\hat{A}^\dagger =  \frac{- \i}{\sqrt{2\hbar}} \left( \bar{P}(t) \, \hat{q} -  \bar{Q}(t) \, \hat{p} \right)\, ,
\end{equation}
where $(Q, P)$ are solutions of the linear system of equations \eqref{Eq-Com-Var}. These operators are actually the linear invariant operators defined in Eq.~\eqref{Inv-bosonic-op}; they are such that
	\beq
	\frac{\d \hat{A}}{\d t} = \i \hbar \, [ \hat{A} , \hat{H} ] 
	+ \frac{\partial \hat{A}}{\partial t} = 0 \, .
	\eeq

Therefore in the most general case, one has the immersion of the constant of motion $\alpha_{\tiny \mbox{Inv}}= \langle \hat{A} \rangle$ into the Hilbert space $ \lag^2(\R, \d q)$ (or  $\lag^2(\R, \d p)$), i.e.
	\beq
	i : \alpha_{\tiny \mbox{Inv}} \mapsto  \psi(\alpha_{\tiny \mbox{Inv}}, q) \, . 
	\eeq
Such an immersion gives rise to the Gaussian wave packet Eq.~\eqref{Gaussian-WP} in the position representation (or \eqref{P-Gaussian-WP} in the momentum representation) as a solution of the Schr\"odinger equation, where
the maximum of the packet follows the classical Hamiltonian equations of motion in Eq. \eqref{Ehrenfest-Teo}, 
whereas the evolution of the uncertainties and their correlations may be obtained by the Hamiltonian equations of motion in Eq.~\eqref{Eq-Com-Var}.

%%%%%%%%%%%%%%%%%%%%%%%%%%%%%%%%%%%%%%%%%%%%%%%%%%%%%%%%%%%%%%%%%%%%%%%%%%%%%%%%%%%%%%%%%%%%%%%%%%%%%%%%%%%%%%%%%%%%%%%%%%%%%%%%%%%%%%%%%%%%%%%%%%%%%%%%%%%%%%%%%%%%%%%%

\section{From Linear to Nonlinear Dynamics}

\label{Section-3}

So far it has been shown that the evolution of the second moments $(\sigma_q, \sigma_p, \sigma_{qp})$ may be determined by means of the solution of the equation of motion \eqref{Eq-Com-Var} under the constraint \eqref{QP-Constrain}.
This section is devoted to show that this evolution is directly connected to a nonlinear one, which has important applications in quantum optics.

\begin{figure}[! h]
	\centering
	\includegraphics[width = 15 cm]{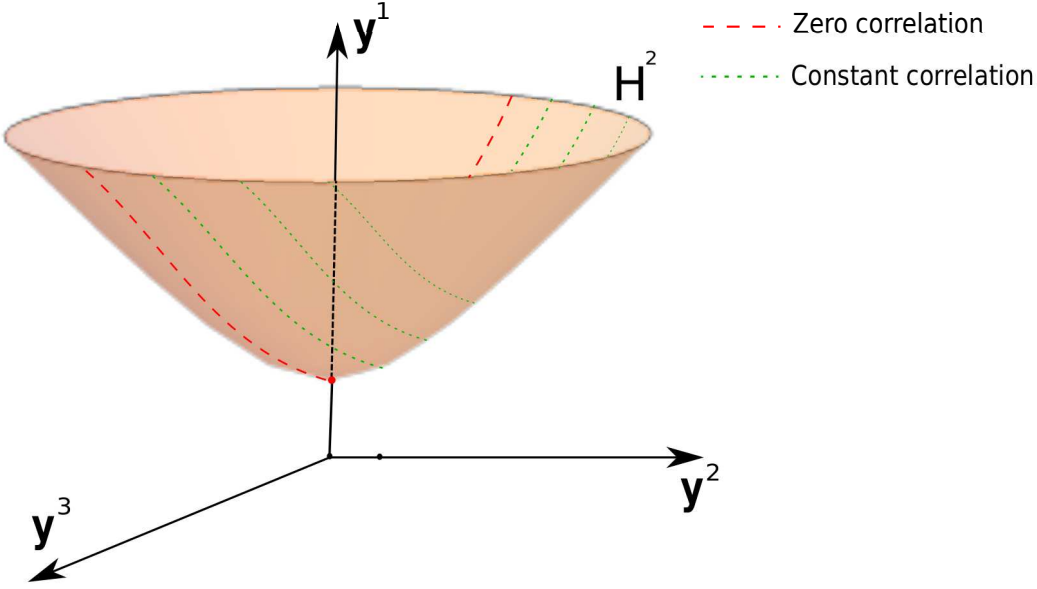}
	\caption{Upper sheet of hyperboloid $\mathbf{H}^2$, where each point on it represents a generalized 
	coherent 	state. In particular, in the dashed red line we have plotted the coherent states and in green dotted
	lines the 	states with constant correlation.}
	\label{Fig-2}
	\end{figure}

Let us start studying the geometrical implications involved in the relation \eqref{Sigma-Com}. 
The action of the Lie algebra $\sl(2,\R)$ in $\R^2$ consists of the matrices $\Sigma$, 
	\beq
	\tilde{\varphi}: \sl(2,\R) \to \mbox{Lin}(\R^2 , \R^2): a \mapsto \Sigma \, ,
	\eeq 
such that $\Tr \, \Sigma = 0$ and $\det \Sigma=1$. 
Then, it is clear that the covariance matrix $\Sigma$, defined in Eq.~\eqref{Sigma-Com}, is an element of $\sl(2,\R)$.
Thus a general element $\Sigma \in \sl(2,\R)$ may be expressed in terms of the basis
	\beq
	\label{sl-basis}
	e_1 = \left(
	\begin{array}{c c}
	0  &  1  \\
	-1 &   0   \\    
	\end{array}
	\right) \, ,
	\quad
	e_2 =\left(
	\begin{array}{c c}
	1  &  0  \\
	0 &   -1   \\    
	\end{array}
	\right) \, ,
	\quad
	e_3 = \left(
	\begin{array}{c c}
	0  &  1  \\
	1 &   0   \\    
	\end{array}
	\right)
	\eeq 
as
	\beq \label{Unc-Sq}
	\Sigma = y^k e_k \, .
	\eeq
This expression directly establishes a one-to-one correspondence between real traceless matrices and vectors in $\R^3$ by
	\beq
	\left(
	\begin{array}{ccc}
	y^2  & y^3 + y^1  \\
	& \\
	y^3 - y^1 & - y^2   \\   
	\end{array}
	\right)
	\mapsto
	(y^1, y^2, y^3)
	\, .
	\eeq
Moreover, the constraint $\det \Sigma = 1$ defines the hyperboloid of two sheets $\mathbf{H}^2 \subset \R^3$ as the manifold\footnote{This description of the quantum system is also known as the ``Pseudosphere''~\cite{balazs1986}.}
	\beq \label{2-Hyperboloid}
	\mathbf{H}^2 = \{ (y^1, y^2, y^3) \in \R^3 \, | \, (y^1)^2 - (y^2)^2 - (y^3)^2 = 1 \} \, .
	\eeq	
Then considering only the upper sheet, see Fig.~\ref{Fig-2}, one may see that each point in this manifold represent a generalized coherent state. 
For example, a coherent state, characterized by $\sigma_{qp} = 0$, corresponds to the points with $y^2 = 0$ of the hyperboloid, i.e. with the points in the curve
	\beq
	\mathbf{H}^1 = \{ (y^1, y^3) \in \R^3 \, | \, (y^1)^2  - (y^3)^2 = 1 \} \, .
	\eeq
This curve is plotted in Fig.~\ref{Fig-2} in a red-dashed line. 
In particular, the point $(1, 0 , 0) \in \mathbf{H}^2$ corresponds to the state with equal uncertainties $\sigma_q = \sigma_p$; then, all points on the hyperboloid different from this minimum characterize a squeezed state.
Another example of interest is to consider states that preserve the correlation between the position and momentum, these are represented in Fig.~\ref{Fig-2} as green-dotted lines, where all of them are curves parallel to $\mathbf{H}^1$.

In consequence the upper sheet of $\mathbf{H}^2$ plays an important role in the description of the \emph{squeezed states}~\cite{scully1999}. To connect with the usual squeezing parameters $(\tau, \varphi)$, we describe the coordinates $y^k$ by means of the Hyperbolic coordinates, this is considering the atlas
	\beq
	\phi : \mathbf{H}^2 \to \R^2 : (y^1, y^2, y^3) \mapsto (\tau, \varphi)
	\eeq 
defined by
	\begin{align}
	y^1 = \cosh{\tau}\, , 
	\quad
	y^2 = \sinh{\tau} \cos{\varphi} \, , 
	\quad
	y^3 = \sinh{\tau} \sin{\varphi} \, .
	\end{align}
	
The system of coordinates $(\tau, \varphi)$ is connected with the \emph{squeezing operator} $\hat{S}$, which allows to immerse $\mathbf{H}^2$ into the Hilbert space $\h$ as follows.
Given the fiducial normalized state $| \, w \rangle $ in the Hilbert space $\h$, one employs the operator $\hat{S}$ to immerse $\mathbf{H}^2$ in the Hilbert space by
	\beq \label{squeeze-immersion}
	s : \xi \in \mathbf{H}^2 \to | \xi \rangle \in \h
	\quad
	\mbox{given by}
	\quad
	| \xi \rangle = \hat{S}(\xi) | w \rangle \, ,
	\eeq
where the coordinate $\xi \in \C$ is defined by $\xi = 2 \, \tau \, \e^{\i \, \varphi}$.
The operator $\hat{S}(\xi)$ is defined as
	\beq
	\hat{S}(\xi) = \e^{ \bar{\xi} \, \hat{K}_{-} - \, \xi \, \hat{K}_{+}}
	\eeq
where the operators $\hat{K}_{\pm}$ are elements of the $\su(1,1)$ algebra, which is defined in terms of the creation and annihilation operators by
	\beq
	\hat{K}_{-} = \frac{1}{2} \, \hat{a}^2 \, ,
	\quad
	\hat{K}_{+} = \frac{1}{2} \, (\hat{a}^\dagger)^2 \, ,
	\quad
	\hat{K}_{0} = \frac{1}{4} \, ( \hat{a} \hat{a}^\dagger + \hat{a}^\dagger \hat{a}) \, ,
	\eeq
with the commutation relations
	\beq
	[\hat{K}_{+}, \hat{K}_{-}] = - 2 \hat{K}_0 \, , 
	\quad
	[\hat{K}_0, \hat{K}_\pm] = \pm \hat{K}_{\pm} \, .
	\eeq

Let us be more explicit in the immersion \eqref{squeeze-immersion}. 
The fiducial normalized state $| w \rangle$ is usually defined by means of the condition $K_{-} | w \rangle = 0$, this condition is satisfied by the Fock states $| 0 \rangle$ and $| 1 \rangle$. 
Considering the Hilbert space of quadratically integrable functions $\lag^2(\R, \d q)$,
the immersed submanifold $s(\mathbf{H}^2) \subset \h$ depends on the choice of the fiducial state; then, considering the state $| 0 \rangle$ the immersed submanifold in $\h$ corresponds to the Gaussian wave packets $\psi(\xi, q) = \langle q | \hat{S}(\xi) | 0 \rangle$. 
These states are known in quantum optics as the \emph{squeezed vacuum states}, whose expectation values are $(\langle \hat{q} \rangle, \langle \hat{p} \rangle) = (0,0)$.
On the other hand, considering the state $| 1 \rangle$ the immersed submanifold clearly does not belong to the space of Gaussian wave functions; therefore, it will not be considered further in this work.

In quantum optics the generalized coherent state is also known as \emph{squeezed coherent state}~\cite{scully1999} and is denoted by $| \alpha, \xi \rangle$ being the result of the following immersion
	\beq \label{Sq-Ch-St}
	h : \mathbf{H}^2 \times \C \to \h 
	\quad
	\text{defined as}
	\quad
	\hat{D}(\alpha) \hat{S}(\xi) | \, 0 \rangle = | \alpha, \xi \rangle \, .
	\eeq
From this definition we see that the parametrization of the generalized coherent state is composed by two different parametrizations. 
The immersion of the complex plane $\C$ parametrizes the expectation values, whereas the immersion of the hyperboloid $\mathbf{H}^2$ parametrizes the second moments. 
For instance, in the immersion defined before, i.e. in Eq. \eqref{Sq-Ch-St}, the action of the operator $\hat{S}(\xi)$ on $| 0 \rangle$ fixes the point on the hyperboloid, later the action of the operator $\hat{D}(\alpha)$ fixes the point in the complex plane.

\vsp

As it has been seen, an important consequence of immersing a ``classical'' submanifold in the Hilbert space is the fact that the time-dependence of the wave function is completely parametrized by the evolution on such a submanifold. 
The manifold $\mathbf{H}^2$ is embedded with the symplectic form
	\begin{align}
	\omega_{\mathbf{H}^2} & = \frac{1}{1 + y^1} \left[
	(1 + y^1) \d y^3 \wedge \d y^2 + y^3 \d y^3 \wedge \d y^2 + y^2 \d y^3 \wedge \d y^2 
	\right]
	 \nonumber \\
	& = \sinh \tau \, \d \varphi \wedge \d \tau  \, .
	\end{align}
Thus there is a Hamiltonian dynamics, $X_{\mathbf{H}^2} \in \X(\mathbf{H}^2)$, which is defined by 
\[
i_{X_{\mathbf{H}^2}} \omega_{\mathbf{H}^2} = - \d e_H  \, .
\]
To construct the Hamiltonian function $e_H \in \F(\mathbf{H}^2)$, let us notice that in quantum mechanics the special linear Lie algebra $\sl(2)$ is given by the operators
	\beq
	\hat{L}_1 = \frac{1}{4} \left( \hat{p}^2 - \hat{q}^2 \right) \, ,
	\quad
	\hat{L}_2 = \frac{1}{4} \left( \hat{q}\,\hat{p} + \hat{p}\,\hat{q} \right) \, ,
	\quad
	\text{and}
	\quad
	\hat{L}_0 = \frac{1}{4} \left( \hat{p}^2 + \hat{q}^2 \right)
	\eeq 
with the commutation relations\footnote{The connection between the elements of $\sl(2, \R)$  with the elements of $\su(1,1)$ is
	\beq
	\hat{K}_{-} = \frac{1}{\hbar} \, ( \hat{L}_1 - \i \, \hat{L}_2 ) \, ,
	\quad
	\hat{K}_{+} = \frac{1}{\hbar} \, ( \hat{L}_1 + \i \, \hat{L}_2) \, ,
	\quad
	\hat{K}_{0} = \frac{1}{\hbar} \, \hat{L}_0 , \nonumber
	\eeq
they are different real-forms of the same complex Lie algebra $\sl(2, \C)$.
	}
	\beq
	[\hat{L}_1, \hat{L}_2] = -\i \, \hbar \, \hat{L}_0 \, , 
	\quad
	[\hat{L}_0, \hat{L}_1] = \i \, \hbar \, \hat{L}_2 \, 
	\quad
	\text{and}
	\quad
	[\hat{L}_0, \hat{L}_2] = -\i \, \hbar \, \hat{L}_1\, .
	\eeq	
In terms of these operators the quadratic Hamiltonian operator \eqref{2-Quad-Ham} has the form
	\beq
	\hat{H} = ( H_2 + H_1 ) \, \hat{L}_0 + 2 V \, \hat{L}_2 + ( H_2 - H_1 ) \, \hat{L}_1 \, ,
	\eeq
and hence the Hamiltonian function on the hyperboloid is simply the expectation value 
	\begin{align}
	e_H & = \langle \, \xi \, | \hat{H} | \, \xi \, \rangle \nonumber \\
	& = ( H_2 + H_1 ) \, y^1 + 2 V \, y^2 + ( H_2 - H_1 ) \, y^3 \nonumber \\
	& = ( H_2 + H_1 ) \, \cosh{\tau} + 2 V \, \sinh{\tau} \cos{\varphi} + ( H_2 - H_1 ) \, \sinh{\tau} \sin{\varphi} \, .
	\end{align}
Finally, from the definition of the symplectic evolution, it is not difficult to find that the Hamiltonian dynamics has the form
	\beq
	X_{\mathbf{H}^2} = - (2 V \sin{\varphi} + (H_1 - H_2)\cos{\varphi}) \frac{\partial }{\partial \tau} 
	- \left[ H_1 + H_2 + ( 2 V \cos{\varphi} + (H_2 - H_1)\sin{\varphi} ) 
	\coth{\tau} \right] \frac{\partial }{\partial \varphi } \, ,
	\eeq
with the Hamiltonian equations of motion
	\begin{align} \label{Hyp-ham-eq}
	\dot{\tau} & = \frac{1}{\sinh{\tau}} \frac{\partial e_H }{\partial \varphi} 
	= - 2 V \sin{\varphi} - (H_1 - H_2)\cos{\varphi} \nonumber \\ 
	\dot{\varphi} & = - \frac{1}{\sinh{\tau}} \frac{\partial e_H }{\partial \tau} 
	= - \left[ 2 V \cos{\varphi} - (H_1 - H_2)\sin{\varphi} \right] \coth{\tau} - (H_1 + H_2)\, ,
	\end{align}
which clearly is a nonlinear dynamics.

\vsp

Now, the connection between the manifold $M$ defined in Eq.~\eqref{M-def} and the hyperboloid $\mathbf{H}^2$ in Eq.~\eqref{2-Hyperboloid} is established. 
To do that, recall that we may represent an element of the $\text{SL}(2, \R)$ group by
	 \beq
	\mathbf{s} = x^\mu e_\mu \, ,
	\eeq
where $\mu = 0, 1, 2, 3$, being $e_0$ the identity matrix and $e_k$ given in \eqref{sl-basis}.
Hence, one may introduce the map 
	\beq
	\left(
	\begin{array}{ccc}
	x^0 + x^2  & x^3 + x^1  \\
	& \\
	x^3 - x^1 & x^0 - x^2   \\     
	\end{array}
	\right)
	\mapsto
	(x^0, x^1, x^2, x^3)
	\, .
	\eeq
Moreover, the coordinates $x^\mu$ define
	\beq \label{Hyp-R4}
	\mathbf{H}^3 = \{ (x^0, x^1, x^2, x^3) \in \R^4 \, | \, (x^0)^2 + (x^1)^2 - (x^2)^2 - (x^3)^2 = 1 \} 
	\eeq
from the constraint $\det \mathbf{s} = 1$. 

Now, expressing the complex coordinates $Q$ and $P$ in terms of their real and imaginary parts, that is, 
 $Q = Q_{\tiny \mbox{R}} + \i \, Q_{\tiny \mbox{I}}$ and $P = P_{\tiny \mbox{R}} + \i \, P_{\tiny \mbox{I}}$, such that
	\begin{align} \label{comp-hyp}
	Q_{\tiny \mbox{R}} & = x^1 - x^3 \, , \qquad Q_{\tiny \mbox{I}} = x^2 - x^0 \, , \nonumber \\
	P_{\tiny \mbox{R}} & = x^2 + x^0 \, , \qquad P_{\tiny \mbox{I}} = x^1 + x^3 \, .
	\end{align}
So, the constraint \eqref{QP-Constrain} in this real variables reproduces the condition in definition \eqref{Hyp-R4}, i.e., one has the transformation $\nu : M \to \mathbf{H}^3$ and the Hamiltonian dynamics $X_{\mathbf{H}^3}$ is defined by $i_{X_{\mathbf{H}^3}} \omega_{\mathbf{H}^3} = - \, \d H_{\mathbf{H}^3}$ with
	\beq
	H_{\mathbf{H}^3} = \nu \circ H_M \, ,
	\quad
	\omega_M = \nu^\ast \omega_{\mathbf{H}^3}
	\quad
	\text{and}
	\quad
	X_{\mathbf{H}^3} = \nu_\ast X_M \, .
	\eeq	
Explicitly the components of the vector field $X_{\mathbf{H}^3}$ define the Hamiltonian equations of motion
	\begin{align} \label{H3-Evo}
	\dot{x}^0 & = - \frac{1}{2}(H_2 + H_1) x^1 - \frac{1}{2}(H_2 - H_1) x^3 - V x^2\, ,   \nonumber \\
	\dot{x}^1 & = \frac{1}{2}(H_2 + H_1) x^0 + \frac{1}{2}(H_2 - H_1) x^4 - V x^3 \, ,   \nonumber \\
	\dot{x}^2 & = \frac{1}{2}(H_2 + H_1) x^3 + \frac{1}{2}(H_2 - H_1) x^1 - V x^0 \, ,  \nonumber \\
	\dot{x}^3 & = - \frac{1}{2}(H_2 + H_1) x^2 - \frac{1}{2}(H_2 - H_1) x^0 - V x^1 \, .
	\end{align}	
Now, to obtain the connection between the hyperboloids $\mathbf{H}^3$ and $\mathbf{H}^2$, notice that
	\beq
	\text{SL}(2, \R) \to  \sl(2, \R) : \mathbf{s} \mapsto \mathbf{s} \, e_1 \, \mathbf{s}^{-1} \equiv y^k e_k  \, ,
	\eeq
where, as one may prove, the matrix  $ \mathbf{s} \, e_1 \, \mathbf{s}^{-1}$ is a traceless matrix.
To show that $y^k e_k$ is an element of $\mathbf{H}^2$, it is enough to observe that
	\beq
	\det (y^k e_k) = \det ( \mathbf{s} \, e_1 \, \mathbf{s}^{-1}) = 1 \, .
	\eeq
To show the way $(y^1, y^2, y^3) \in \mathbf{H}^2$ depends on $(x^0, x^1, x^2, x^3) \in \mathbf{H}^3$, one simply uses the explicit form of the matrices $\mathbf{s}$ and $e_1$ to obtain
	\begin{align} \label{Hyp-coord}
	y^1 & =   (x^0)^2 + (x^1)^2 +(x^2)^2 + (x^3)^2 \, , \nonumber \\
	y^2 & =   2 \, (x^1 x^2 - x^0 x^3) \, , \nonumber \\
	y^3 & =  2 \, (x^1 x^3 + x^0 x^2) \, .
	\end{align}	
Then one arrives at the covering map
	\beq \label{H3-to-H2}
	\chi : \mathbf{H}^3 \to \mathbf{H}^2 : (x^0, x^1, x^2, x^3) \mapsto (y^1, y^2, y^3) \, .
	\eeq

Note that the generalization of the results in this sections for more degrees of freedom is not simple, because the connection between the Lie group $\text{SL}(2n, \R)$ and its Lie algebra $\sl (2n, \R)$, for $n > 1$, is more complex.

%%%%%%%%%%%%%%%%%%%%%%%%%%%%%%%%%%%%%%%%%%%%%%%%%%%%%%%%%%%%%%%%%%%%%%%%%%%%%%%%%%%%%%%%%%%%%%%%%%%%%%%%%%%%%%%%%%%%%%%%%%%%%%%%%%%%%%%%%%%%%%%%%%%%%%%%%%%%%%%%%%%%%%%%

\section{Nonlinear Riccati Evolution}

\label{Section-4}

In the last section, we have shown that for the one-dimensional parametric oscillator system it is possible to associate a nonlinear dynamics with the evolution of the Gaussian wave packets; however, this nonlinear description appears to be more difficult for more degrees of freedom.
Here we will show that there is another nonlinear description of the quadratic Hamiltonian systems independent of the degrees of freedom of the system; such a nonlinear description is the Riccati evolution.
The nonlinear Riccati evolution has currently gained considerable interest and has been widely studied and applied to quantum systems~\cite{cruz_schuch_castanos_rosas-ortiz-1, cruz_schuch_castanos_rosas-ortiz-2, schuch2018}.

Let us introduce the Riccati evolution for one-dimensional quantum systems by means of the transformation
	\beq \label{Ric-Q}
	\pi : M \to \mathbb{HP}^2 : (Q, P) \mapsto \Ri = \frac{P}{Q} \, ,
	\eeq
where the manifold $M$ has been defined in \eqref{M-def}, whereas the space $\mathbb{HP}^2$ is known as the \emph{Siegel upper half plane} \cite{siegel1943, mcduff2017} and is defined as
	\beq \label{Siegel-plane}
	\HP^2 = \{ \Ri \in \C \, | \, \Ri_{\tiny \mbox{I}} > 0 \} \, ,
	\eeq 
considering $\Ri = \Ri_{\tiny \mbox{R}} + \i \, \Ri_{\tiny \mbox{I}}$. 
Then, by means of the linear equations \eqref{Eq-Com-Var} follows that the dynamics on the manifold $M$ induces a nonlinear dynamics in the space $\HP^2$ such that the integral curves of this dynamics are solutions of the Riccati equation (associated with the Hamiltonian defined in Eq.~(\ref{2-Quad-Ham}))
	\beq \label{Osc-Ric-Q}
	\dot{\Ri} + H_2 \, \Ri^2  + 2 V \, \Ri + H_1 = 0 \, .
	\eeq
In fact, one may express the Gaussian wave packet \eqref{Gaussian-WP} in terms of the Riccati variables $\Ri \in \HP^2$ as\footnote{Remember that the parameter $\alpha$ is given in terms of the expectation values of $\hat{p}$ and $\hat{q}$, see Eq.~(\ref{alpha-coordinates}).}
	\beq \label{G-WP-Ri-q}
	\psi(\alpha, \Ri, q) = \frac{1}{ (\pi \, \hbar)^{1/4}}
	\exp
	\left\{
	\frac{\i}{2 \hbar} \, \Ri \, ( q - \langle \hat{q} \rangle)^2 
	+ \frac{\i}{\hbar} \langle \hat{p} \rangle ( q - \langle \hat{q} \rangle) 
	+ \frac{\i}{2 \hbar} \,  \langle \hat{q} \rangle \,  \langle \hat{p} \rangle 
	-\frac{1}{2} \int^t [H_2 \, \Ri(t') + V ]\d t'
	\right\} \, .
	\eeq
On the other hand, it is also posible to consider the coordinates
	\beq \label{Ric-P}
	\tilde{\pi} : M \to \mathbb{HP} : (Q, P) \mapsto \tilde{\Ri} = \frac{Q}{P} \, ,
	\eeq
where for this case we may express the generalized coherent state in the momentum representation as
	\beq 
	\tilde{\psi}(\alpha, \tilde\Ri, p) = 
	\frac{\sqrt{ \i }}{ (\pi \, \hbar)^{1/4}} 
	\exp
	\left\{
	- \frac{\i}{2 \hbar} \, \tilde{\Ri} \, ( p - \langle \hat{p} \rangle)^2 
	- \frac{\i}{\hbar} \langle \hat{q} \rangle ( p - \langle \hat{p} \rangle) 
	- \frac{\i}{2 \hbar} \,  \langle \hat{q} \rangle \,  \langle \hat{p} \rangle 
	+ \frac{1}{2} \int^t [H_1 \, \tilde{\Ri}(t') + V ]\d t'
	\right\} \, ,
	\eeq
where now the time dependent complex function $\tilde{\mathcal{C}}(t)$ obeys the nonlinear Riccati equation 
	\beq \label{Osc-Ric-P}
	- \dot{\tilde \Ri} + H_1 \, \tilde{\Ri}^2 + 2 V \, \tilde{\Ri} + H_2 = 0 \, .
	\eeq
Therefore, these nonlinear Riccati equations are closely connected with the evolution of the generalized coherent states.	

The geometrical nature of the transformations \eqref{Ric-Q} and \eqref{Ric-P} is the following.
As one can see the linear system of equations \eqref{Eq-Com-Var} together with its constraint~\eqref{QP-Constrain} is invariant under the multiplication by a global phase factor, i.e. it is invariant under the tranformation 
	\beq
	(Q, P) \mapsto \e^{i \phi}(Q, P) \, .
	\eeq
This invariance allows to reduce the dynamics into a manifold of a lower dimension for the dynamical system, for a complete review about the reduction procedure see Ref.~\cite{marmo1985}.
The multiplication by the global phase factor is the action of the group $U(1)$, which may be described infinitesimally by means of the linear vector field
	\beq
	\Gamma = P_{\tiny \mbox{R}} \, \frac{\partial}{\partial Q_{\tiny \mbox{R}}} 
	- Q_{\tiny \mbox{R}} \, \frac{\partial}{\partial P_{\tiny \mbox{R}}} 
	+ P_{\tiny \mbox{I}} \, \frac{\partial}{\partial Q_{\tiny \mbox{I}}} 
	- Q_{\tiny \mbox{I}} \, \frac{\partial}{\partial P_{\tiny \mbox{I}}} \, .
	\eeq
Then, on the manifold $M$ there is defined a regular distribution $\mathcal{D} = \{ \Gamma \}$, whose integral curves 
	\beq
	\ell_{(r_1,r_2)} = \{ (Q_{\tiny \mbox{R}}, P_{\tiny \mbox{R}}, Q_{\tiny \mbox{I}}, P_{\tiny \mbox{I}}) \in \R^4 \, | \, 
	 P^2_{\tiny \mbox{R}} + Q^2_{\tiny \mbox{R}} = r^2_1 \cup 
	 P^2_{\tiny \mbox{I}} + Q^2_{\tiny \mbox{I}} = r^2_2 \,
	 \} 
	\eeq
are a family $\{ \ell_{(r_1,r_2)} \}$ of disjoint subsets which foliates the manifold $M$, with $r_1$ and $r_2$ real constant quantities.
Furthermore, the foliation defines the equivalence relation $\Phi^{\Gamma}$ as
	\beq
	(Q_{\tiny \mbox{R}}, P_{\tiny \mbox{R}}, Q_{\tiny \mbox{I}}, P_{\tiny \mbox{I}}) \in \Phi^{\Gamma}
	\quad
	\text{iff}
	\quad
	(Q_{\tiny \mbox{R}}, P_{\tiny \mbox{R}}, Q_{\tiny \mbox{I}}, P_{\tiny \mbox{I}}) \in \ell_{(r_1,r_2)} \  .
	\eeq 
Thus, it is possible to define the quotient space $M / \Phi^{\Gamma}$ with respect to the equivalence relation defined by the foliation, which is identified with the \emph{Siegel upper half plane} $\HP^2$~\cite{siegel1943, mcduff2017}.

Now that the canonical projection $\pi$ from the manifold $M$ to the Siegel upper half plane $\HP^2$ has been established, one may apply this results to the dynamics of the system.
Then, the dynamical vector field $X_M \in \X(M)$, given in \eqref{Dynamics-M}, is \emph{projectable}
onto a dynamics $X_{\HP^2}  \in \X(\HP^2)$, because	
	\beq
	[X_M, \Gamma] = 0 \, ,
	\eeq
for a formal proof of this result see reference~\cite{marmo1985}.
Then, because the dynamics is projectable, it carries leaves of the foliation $\Phi^{\Gamma}$, into leaves, i.e. the foliation is invariant under $X_M$~\cite{marmo1985}. 
In this sense the group action of $U(1)$ is a symmetry for the dynamics.
Therefore, $\HP^2$ is the space that results after taking into account the symmetry of the multiplication by a global phase factor in the linear equation of motion \eqref{Eq-Com-Var}, which gives rise to a nonlinear evolution.

\vsp

\begin{figure}[! t]
	\centering
	\includegraphics[width = 16 cm]{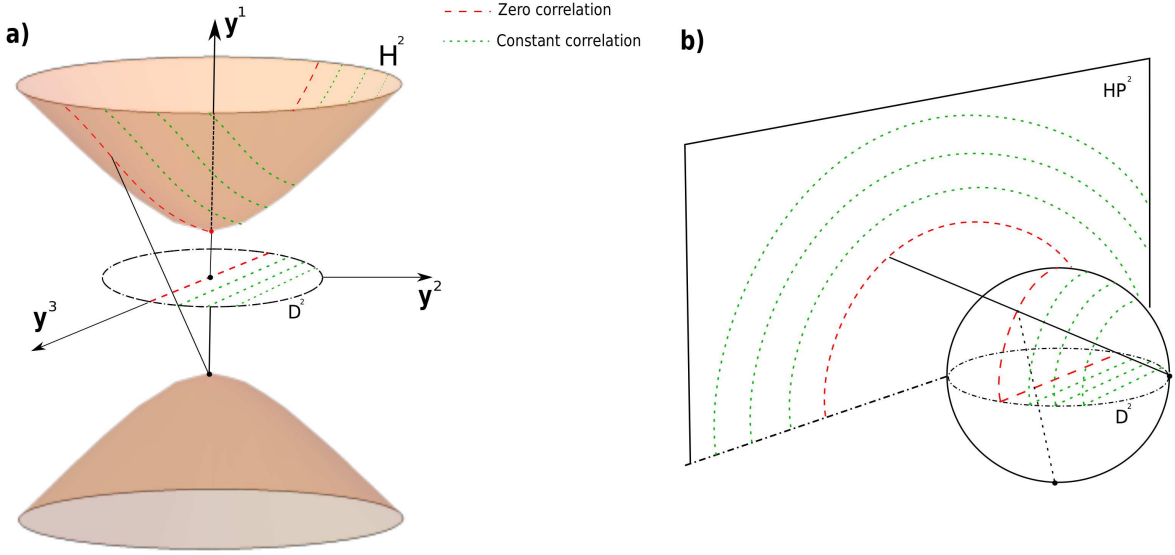}
	\caption{a) Projection of the upper sheet of the hyperboloid $\mathbf{H}^2$ onto the Poincar\'e disk 
	$\mathbb{D}^2$ with the projection point $(-1,0,0)$. b) Stereographic projection of the Poincar\'e disk onto the 	``north hemisphere''  of the sphere, later the hemisphere points are projected onto the Siegel upper half plane 	$\HP^2$.}
	\label{Fig-3}
	\end{figure}

To have a closed picture of our study, now the connection between the upper sheet of the hyperboloid  with the Siegel upper half plane is established.
So, it is well-known that every point on the hyperboloid $\mathbf{H^2}$, defined in Eq.~\eqref{2-Hyperboloid}, may be projectable onto the \emph{Poincar\'e disk} or in the Siegel upper half plane. 
In the first case, one considers the projection point $(-1, 0, 0) \in \R^3$, such that a point $(y^1, y^2, y^3) \in \mathbf{H}^2$ is projected onto the plane $y^1 = 0$, see Fig.~\ref{Fig-3}a, given by the map
	\beq \label{Disk-Projection}
	\zeta = \frac{y^2 + \i \, y^3}{1 + y^1} \, \, ,
	\eeq
i.e., the result is the projection
	\beq
	v : \mathbf{H}^2 \to \mathbb{D}^2 : (y^1, y^2, y^3) \in \R^3 \mapsto \zeta \in \C \, ,
	\eeq
where the complex number $\zeta = \zeta_{\tiny\mbox{R}} + \i \, \zeta_{\tiny\mbox{I}}$ is an element of the Poincar\'e disk defined as the open disk
	\beq
	\mathbb{D}^2 = \{ \zeta \in \C \, | \, | \zeta | < 1 \}\, .
	\eeq
This projection is displayed in Fig.~\ref{Fig-3}a, where in addition it is possible to see that the curves for zero correlation and constant correlation are mapped on parallel lines on the Poincar\'e disk.
Notice that the evolution in the hyperboloid has induced a nonlinear evolution in $\mathbb{D}^2$ given by the Riccati equation
	\beq  \label{Riccati-Disk}
	\dot{\zeta} - \frac{1}{2}(H_1 - H_2 - 2 \, \i \, V )\zeta^2 + \i \, (H_2 + H_1) \zeta 
	+ \frac{1}{2}(H_2 - H_1 + 2 \, \i \, V ) = 0 \, .
	\eeq 
	
Finally, it is possible to obtain the upper half Siegel plane from the Poincar\'e disk as follows.
First, we consider the stereographic projection of the Poincar\'e disk onto the``north hemisphere'' of a sphere, employing as a projection point the ``south pole'' of the sphere, see Fig~\ref{Fig-3}b. 
Later, the hemisphere points are projected onto the tangent plane to the sphere by means of another stereographic projection, see Fig.~\ref{Fig-3}b. 
In Fig.~\ref{Fig-3}b it is plotted the projection on the sphere and in $\HP^2$ the curves for $\sigma_{qp} = 0$ (red lines) and $\sigma_{qp} = \text{cte}$ (green lines).

Therefore, the open disk $\mathbb{D}^2$ is mapped bijectively to the half plane $\mathbb{HP}^2$ by the map
	\beq
	u : \mathbb{D}^2 \to \mathbb{HP}^2 : \zeta \in \C \mapsto \Ri \in \C
	\eeq
by means of the M\"obious transformation 
	\beq \label{bij-D-to-HP} 
	\Ri = \frac{\zeta + \i}{\i \zeta +1}\, .
	\eeq
Consequently we may construct the map $u \circ v : \mathbf{H}^2 \to \HP^2$ to obtain the relations 
	\beq
	y^1 = \frac{1 + \Ri \bar{\Ri}}{ 2 \, \Ri_{\tiny\mbox{I}}}  \, ,
	\qquad
	y^2 = \frac{ \Ri_{\tiny\mbox{R}} } { \Ri_{\tiny\mbox{I}} } \, ,
	\qquad
	y^3 =\frac{ - 1+ \Ri \bar{\Ri} } {2 \, \Ri_{\tiny\mbox{I}}} \, ,
	\eeq
or directly from the uncertainties and correlation of position and momentum, one has that
	\beq \label{Ricc-Eq-Com}
	\Ri = \frac{\sigma_{qp}}{\sigma^2_q} + \frac{\i \, \hbar}{2} \, \frac{1}{\sigma_q^2} \, .
	\eeq
We have obtained two new different immersions for the generalized coherent states
	\beq
	d : \mathbb{D}^2 \times \C \to \h : (\zeta, \alpha) \mapsto | \zeta, \alpha \rangle
	\quad
	\text{and}
	\quad
	g : \HP^2 \times \C \to \h : (\Ri, \alpha) \mapsto | \Ri, \alpha \rangle \, .
	\eeq
	
Finally, we may summarize all our connections in the following diagramme
	
	\begin{diagram}
	M		& 	\rTo_{\nu} 	&	\mathbf{H}^3     \\
			&			&      \dTo_{\chi}	      	 \\
	\dTo_{\pi} &			&	\mathbf{H}^2	\\ 
			&			&     	\dTo_{v}		 \\
	\HP^2	&	\lTo_{u}	& 	\mathbb{D}^2
	\end{diagram}

After the kinematical picture has been completed, the dynamical properties of the Riccati descriptions are now the subject of interest.
The dynamics on $\HP^2$ is non-linear and we show it is Hamiltonian.
To see this, notice that the upper half plane is endowed with the symplectic form $\omega_{\tiny \mbox{WP}}$ (the Weil--Petersson symplectic form), defined in coordinates as
	\begin{align} \label{WP-form}
	\omega_{\tiny \mbox{WP}} & = \frac{ - 2 \, \i}{(\Ri - \bar{\Ri})^2} \,\, \d \, \bar{\Ri} \wedge \d \,\Ri \nonumber \\
	& = \frac{1}{\Ri_{\tiny \mbox{I}}^2} \, \d \Ri_{\tiny \mbox{I}} \wedge \d \Ri_{\tiny \mbox{R}} \, ,
	\end{align}
considering $\Ri = \Ri_{\tiny \mbox{R}} + \i \, \Ri_{\tiny \mbox{I}}$.
Furthermore, by means of the map in \eqref{Ric-Q}, we may look at the pullback of $\omega_M$ in Eq.~\eqref{Sym-Comp} to $\C^2$ and see that $\omega_M = \pi^\ast \omega_{\tiny \mbox{WP}}$.
Hence, with the help of this symplectic form the evolution in $\mathbb{HP}^2$ is given by the Hamiltonian vector field $X_{\HP}$ defined intrinsically by
	\beq\label{Ham-Dyn}
	i_{X_{\HP}} \omega_{\tiny \mbox{WP}} = - \, \d \, H_{\HP} \, ,
	\eeq	
where the Hamiltonian function in the coordinates $\Ri \in \HP^2$ has the form
	\beq \label{UP-Ham}
	H_{\HP} 
	 = 
	\frac{2 \, \i }{\Ri - \bar{\Ri} } 
	\, \,
	( \,  1  \,\,\, , \,\,\, \bar{\Ri} \, )
	\left(
	\begin{array}{cc}
	H_1  &  V   \\
	& \\
	V &  H_2   \\   
	\end{array}
	\right)
	\left(
	\begin{array}{c}
	 1   \\
	 \\
	 \Ri
	\end{array}
	\right) \, . 
	\eeq
Using the symplectic form \eqref{WP-form} and the Hamiltonian function in \eqref{UP-Ham}, it is straightforward to obtain the Hamiltonian vector field in the explicit form
	\beq \label{Ham-Vec-Field}
	X_{\HP} = X_\Ri \, \frac{\partial}{\partial \Ri} + X_{\bar \Ri} \, \frac{\partial}{\partial \bar{\Ri}} \, ,
	\eeq
where $X_\Ri$ is the complex conjugated of $X_{\bar{\Ri}}$ and it is given by 
	\beq
	X_\Ri = - H_2 \, \Ri^2 - 2\, V \, \Ri - H_1 \, .
	\eeq
Hence, the integral curves of this Hamiltonian vector field are given by the solutions of the Hamiltonian equations of motion
	\begin{align} \label{Ricc-Poin-Eq}
	\dot{\Ri} & = - \frac{(\Ri - \bar{\Ri})^2}{2 \, \i} \, \frac{\partial H}{\partial \bar{\Ri}} 
	= - H_2 \, \Ri^2 - 2\, V \, \Ri - H_1\,, \nonumber \\
	\dot{\bar{\Ri}} & =  \frac{(\Ri - \bar{\Ri})^2}{2 \, \i} \, \frac{\partial H}{\partial \Ri} 
	= - H_2 \, \bar{\Ri}^2 - 2\, V \, \bar{\Ri} - H_1 \, .  
	\end{align}
These equations are identical with the Riccati equation presented in Eq.~\eqref{Osc-Ric-Q} as it should be.
Moreover, we may introduce the Poisson brackets for the upper Siegel half plane space.
So, given the real functions $A , B \in \F(\HP^2)$ the Poisson bracket is given by
	\begin{align} \label{Poiss-Bra}
	\{ A , B \}_{\tiny \mbox{WP}} & = \frac{(\Ri - \bar{\Ri})^2}{2 \, \i}  
	\left[
	\frac{\partial A}{\partial \Ri} \, \frac{\partial B}{\partial \bar{\Ri}} 
	- \frac{\partial A}{\partial \bar{\Ri}} \, \frac{\partial B}{\partial \Ri}
	\right] \nonumber \\
	& = \Ri^2_{\tiny \mbox{I}} \left[
	\frac{\partial A}{\partial \Ri_{\tiny \mbox{I}}} \, \frac{\partial B}{\partial \Ri_{\tiny \mbox{R}}} 
	- \frac{\partial A}{\partial \Ri_{\tiny \mbox{R}}} \, \frac{\partial B}{\partial \Ri_{\tiny \mbox{I}}}
	\right] \, .  
	\end{align}
In particular, the time evolution of any (not explicitly time-dependent) real function on the Siegel half plane $A \in \F(\mathbb{HP}^2)$ is given by
	\beq \label{Ev-HP}
	\frac{\d A}{\d t} = \{ H, A \}_{\tiny \mbox{WP}} \, .
	\eeq

As it has been pointed out, one may also describe the evolution on the Poincar\'e disk, with a Riccati evolution. 
Furthermore, we now are able to prove that the dynamics is also Hamiltonian, where the symplectic structure in $\mathbb{D}^2$ is obtained via the pullback $u^\ast (\omega_{\tiny \mbox{WP}}) = \omega_{\mathbb{D}}$, which in coordinates has the form 
	\begin{align}
	\omega_{\mathbb{D}} & = \frac{2 \, \i}{(1- \zeta \bar{\zeta} )^2 } \, \d \bar{\zeta} \wedge \d \zeta \nonumber \\
	& =  \frac{4}{(1- \zeta \bar{\zeta} )^2 } \, \d \zeta_{\tiny \mbox{I}} \wedge \d \zeta_{\tiny \mbox{R}} \, ,
	\end{align}
such that the Hamiltonian dynamics given by $i_{X} \omega_{\mathbb{D}}  = - \d H_{\mathbb{D}}$ has the Hamiltonian function $H_{\mathbb{D}} = u^{-1} (H_{\HP})$, i.e.,
	\beq \label{U-Ham}
	H_{\mathbb{D}}
	 = 
	\frac{ 1 }{1 - \bar{\zeta} \zeta } 
	\, \, 
	( \,  -\i \bar{\zeta} +1  \,\,\, , \,\,\, \bar{\zeta} - \i \, )
	\left(
	\begin{array}{cc}
	H_1  &  V   \\
	& \\
	V &  H_2   \\   
	\end{array}
	\right)
	\left(
	\begin{array}{c}
	 \i \zeta +1   \\
	 \\
	 \zeta + \i
	\end{array}
	\right) \, . 
	\eeq
Then the Hamiltonian equations of motion are given by
	\begin{align} \label{Ricc-Poincare-Disk}
	\dot{\zeta} & = - \frac{(1 - \zeta \bar{\zeta})^2}{2 \, \i} \, \frac{\partial H_{\mathbb{D}}}{\partial \bar{\zeta}} 
	=  \frac{1}{2}(H_1 - H_2 - 2 \, \i \, V )\zeta^2 - \i \, (H_2 + H_1) \zeta 
	- \frac{1}{2}(H_2 - H_1 + 2 \, \i \, V )  \,, \nonumber \\
	\dot{\bar{\zeta}} & =  \frac{(1 - \zeta \bar{\zeta})2}{2 \, \i} \, \frac{\partial H_{\mathbb{D}}}{\partial \zeta} 
	=  \frac{1}{2}(H_1 - H_2 + 2 \, \i \, V )\bar{\zeta}^2 + \i \, (H_2 + H_1) \bar{\zeta} 
	- \frac{1}{2}(H_2 - H_1 - 2 \, \i \, V ) \, , 
	\end{align}
and hence the evolution of an arbitrary time-independent real function $F \in \F(\mathbb{D}^2)$ is given by
	\beq \label{Ev-HP}
	\frac{\d F}{\d t} = X_{\mathbb{D}} [F] = \{ H, F \}_{\mathbb{D}^2} \, ,
	\eeq
where $\{ \, \cdot \, , \, \cdot \, \}_{\mathbb{D}}$ is the Poisson bracket defined as
	\begin{align} \label{Poiss-Bra}
	\{ H , F \}_{\mathbb{D}} & = \frac{(1 - \zeta \bar{\zeta})^2}{2 \, \i}  
	\left[
	\frac{\partial H}{\partial \zeta} \, \frac{\partial F}{\partial \bar{\zeta}} 
	- \frac{\partial H}{\partial \bar{\zeta}} \, \frac{\partial F}{\partial \zeta}
	\right] \nonumber \\
	& = \frac{(1 - \zeta \bar{\zeta})^2}{4}  
	\left[
	\frac{\partial H}{\partial \zeta_{\tiny \mbox{R}}} \, \frac{\partial F}{\partial \zeta_{\tiny \mbox{I}}} 
	- \frac{\partial H}{\partial \zeta_{\tiny \mbox{I}}} \, \frac{\partial F}{\partial \zeta_{\tiny \mbox{R}}}
	\right] \, .  
	\end{align}

%%%%%%%%%%%%%%%%%%%%%%%%%%%%%%%%%%%%%%%%%%%%%%%%%%%%%%%%%%%%%%%%%%%%%%%%%%%%%%%%%%%%%%%%%%%%%%%%%%%%%%%%%%%%%%%%%%%%%%%%%%%%%%%%%%%%%%%%%%%%%%%%%%%%%%%%%%%%%%%%%%%%%%%%

\section{Degenerate parametric amplification}

\label{Section-5}

In order to apply all the formalism previously developed to a concrete physical system, in this Section the study of the \emph{degenerate parametric amplifier}~\cite{scully1999} is considered . 
In quantum optics the parametric amplifier is an optical device in which there is a coupling of three modes of the electromagnetic field in a nonlinear optical crystal. 
The frequencies involved are $\omega_{\tiny \mbox{p}}$ (pump), $\omega_{\tiny \mbox{i}}$ (idler) and $\omega_{\tiny \mbox{s}}$ (signal) and they are such that $\omega_{\tiny \mbox{p}} = \omega_{\tiny \mbox{s}} + \omega_{\tiny \mbox{i}}$, where for the case of degenerate parametric amplification the idler and the signal frequencies coincide, i.e. $\omega = \omega_{\tiny \mbox{s}} = \omega_{\tiny \mbox{i}}$ and then $\omega_{\tiny \mbox{p}} = 2 \, \omega$. 
The Hamiltonian describing this device is
	\beq 
	\label{Deg-Par-Ampl}
	\hat{H} = \hbar \, \omega \, \hat{a}^\dagger \hat{a} + 2 \, \hbar \, \omega \, \hat{b}^\dagger \hat{b}
	+ \hbar \, \kappa \, [ \, (\hat{a}^\dagger)^2 \, \hat{b} + \hat{a}^2 \, \hat{b}^\dagger \, ] \, .
	\eeq
Here $\hat{b}$ and $\hat{a}$ are the annihilation operators for the pump and the signal (idler) modes, respectively, $\kappa$ is the coupling constant that depends on the properties of the nonlinear crystal. 
We are interested in the \emph{parametric approximation}, i.e. when the pump field is treated as a classical field~\footnote{The interaction of different radiation modes through nonlinear crystals allows the generation of interesting states of light. Most of the theoretical analysis refers to situations where one mode is placed in a high amplitude coherent state. This is called the \emph{parametric approximation}.}. Therefore
the expectation value of the Hamiltonian (\ref{Deg-Par-Ampl}) with respect the coherent state $| \beta \, \e^{- \i \omega t} \rangle$ of the pump field is considered, such that $ \hat{b} \, | \beta \, \e^{- \i \omega t} \rangle = \beta \, \e^{- \i \omega t} | \beta \, \e^{- \i \omega t} \rangle$. 
Thus one has the effective Hamiltonian  
	\begin{align} \label{Eff-Hamiltonian}
	\hat{H}_{\tiny \mbox{Eff}} & = \frac{\hbar \, \omega}{2} \, [ \hat{a}^\dagger \hat{a}  + \hat{a} \hat{a}^\dagger ] 
	+ \hbar \, \kappa \, [ \, (\hat{a}^\dagger)^2 \, \beta \, \e^{- \i \omega t} 
	+ \hat{a}^2 \, \bar{\beta} \, \e^{ \i \omega t} \, ] \nonumber \\
	& = \frac{\hbar}{4} \,
	( \, \hat{a} \,\,\, , \,\,\, \hat{a}^\dagger \, )
	\left(
	\begin{array}{ccc}
	\bar{\xi} \, \e^{ \i \omega t}  & 2 \, \omega  \\
	& \\
	2 \, \omega  & \xi \, \e^{ - \i \omega t}   \\    
	\end{array}
	\right)
	\left( 
	\begin{array}{c}
	\hat{a} \\
	 \\
	\hat{a}^\dagger
	\end{array}
	\right) \, ,
	\end{align}
where a constant term has been ignored and one defines $\xi = 4 \, \kappa \, \beta$.
Then, the dynamics of the complex parameter $\alpha = \langle \alpha | \hat{a} | \alpha \rangle$ is given by the Hamiltonian equations of motion \eqref{alpha-Ham}, which, for our case of interest, have the form
	\begin{align}
	\dot{\alpha} & = - \frac{\i}{2} ( \xi \, \e^{- \i \omega t} \bar{\alpha} + 2 \omega \alpha ) \, , \nonumber \\
	\dot{\bar \alpha} & = \frac{\i}{2} ( \bar{\xi} \, \e^{\i \omega t} \alpha + 2 \omega \bar{\alpha} ) \, .
	\end{align}
The integral curves of this system of differential equations are given by
	\beq \label{Alpha-sol}
	\alpha(t) =
	\begin{cases}
	 \left(\frac{1}{\Omega} [ 2 \dot{\alpha}_0 + \i \, \omega \, \alpha_0 ] \sin{\frac{\Omega t}{2}} 
	 + \alpha_0 \cos{\frac{\Omega t}{2}} \right)\e^{- \frac{\i}{2}\omega t} 
	 & \text{for}  \quad \Omega = \sqrt{\omega^2 - |\xi|^2} \, , \\
	 \\
	  \left(\left[ \dot{\alpha}_0 + \frac{\i \, \omega}{2} \alpha_0 \right] t + \alpha_0 \right)\e^{- \frac{\i}{2}\omega t}
	 & \text{for}  \quad \Omega = 0 \, , \\
	 \\
	  \frac{1}{\tilde{\Omega}}\left([2 \dot{\alpha}_0 + \i \, \omega \, \alpha_0] \sinh{ \frac{\tilde{\Omega} t}{2}} 
	  + \alpha_0 \cosh{ \frac{\tilde{\Omega} t}{2} } \right)\e^{- \frac{\i}{2}\omega t}
	 & \text{for}  \quad \tilde{\Omega} = \sqrt{|\xi|^2 - \omega^2} \, ,
	 \end{cases}
	\eeq
with  the initial conditions $\alpha_0$ and $\dot{\alpha}_0$.	
Thus, the expectation value of the quadratures of the field $(\hat{q}, \hat{p})$ can be obtained by the expression 
	\beq
	\alpha = \frac{1}{\sqrt{2 \hbar}} \left(
	\sqrt{\omega} \,\langle q \rangle + \frac{\i}{\sqrt{\omega}} \langle p \rangle
	\right) \, .
	\eeq
The immersion of the complex number $\alpha$, given in Eq. \eqref{Alpha-sol}, into the Hilbert space $\lag^2(\R, \d q)$ gives rise to the evolution of the coherent states with constant second moments, namely
	\beq
	\sigma_q = \frac{\hbar \omega}{2}  \, ,
	\quad 
	\sigma_p = \frac{\hbar}{2  \omega}
	\quad
	\text{and}
	\quad
	\sigma_{qp} = 0 \, .
	\eeq

Because the aim is not to solve this system in all its generality, let us study the integral curves $\alpha(t)$ in the complex plane only for the first case, $\Omega = \sqrt{\omega^2 - |\xi|^2}$.
So, we may express the solution as
	\beq
	\alpha(t) = 
	\left( 
	\frac{1}{2} \left[ 1 + \frac{\omega}{\Omega} \right] \alpha_0 - \frac{\i \, \dot{\alpha}_0}{\Omega}
	\right) \e^{\frac{\i}{2}(\Omega - \omega) t}
	+
	\left( 
	\frac{1}{2} \left[ 1 - \frac{\omega}{\Omega} \right] \alpha_0 + \frac{\i \, \dot{\alpha}_0}{\Omega}
	\right) \e^{-\frac{\i}{2}(\Omega + \omega) t} \, ,
	\eeq
and, considering the polar forms	
	\beq
	\frac{1}{2} \left[ 1 + \frac{\omega}{\Omega} \right] \alpha_0 - \frac{\i \, \dot{\alpha}_0}{\Omega} 
	= r_1 e^{\i \varphi_1}
	\quad
	\text{and}
	\quad
	\frac{1}{2} \left[ 1 - \frac{\omega}{\Omega} \right] \alpha_0 + \frac{\i \, \dot{\alpha}_0}{\Omega}
	= r_2 e^{\i \varphi_2} \, ,
	\eeq	
one arrives at the final form	
	\beq \label{Alpha-cycloid-form}
	\alpha(t) = r_1 \e^{\frac{\i}{2}(\Omega - \omega)(t + \frac{2 \varphi_1}{\Omega - \omega} )}
	+  r_2 \e^{- \frac{\i}{2}(\Omega + \omega)(t - \frac{2 \varphi_2}{\Omega + \omega} )} \, .
	\eeq

\begin{figure}[! t]
	\centering
	\includegraphics[width = 14 cm]{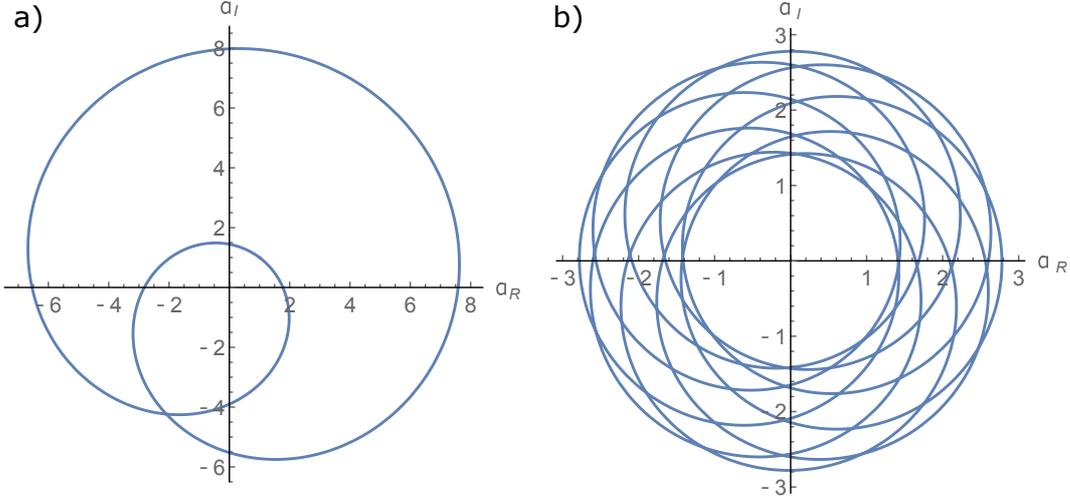}
	\caption{Time evolution in the complex plane of solutions $\alpha(t)$ with initial conditions 
	$\alpha_0 = 1 + \i$ and parameters a) $(\omega, |\xi| ,\Omega) = (6, 4\sqrt{2}, 2)$ and 
	b) $(\omega, |\xi| ,\Omega) = (5, 3, 4)$.}
	\label{Fig-4}
	\end{figure}

On the other hand, in complex calculus the curve defined by a parametrization of the form
	\beq\label{Cycloid-eq}
	C(\theta) = (a + b) \e^{\i \, \mu (\theta - \theta_1)} + d \e^{\i \, \nu (\theta - \theta_2)}\, ,
	\eeq
where $\theta_1$, $\theta_2$, $a$, $b$ and $d$ are real constants, is well-known. If the real
If the real constants  $\mu$ and $\nu$ are such that $\mu \, \nu > 0$, these curves are denoted as \emph{epicycloids} if $|d| = |b|$ and \emph{epitrochoids}
 for any other case. 
Geometrically this curves are traced by a point attached to a circle of radius $b$ rolling around the outside of a fixed circle of radius $a$, where the point is at a distance $d$ from the center of the exterior circle.
Moreover, the curves defined before are closed and periodic iff the quotient $\nu/\mu \in \mathbb{Q}$ is a rational number. 
Comparing the expression of the integral curves in Eq.~\eqref{Alpha-cycloid-form} with the curves in Eq.~\eqref{Cycloid-eq}, it is straightforward that the curves $\alpha(t)$ are epicycloids or epitrochoids in the complex plane with the identification
	\beq
	a + b = r_1 \, ,
	\quad
	d = r_2 \, ,
	\quad
	\theta_1 = - \frac{2 \varphi_1}{\Omega - \omega} \, ,
	\quad
	\theta_2 = \frac{2 \varphi_2}{\Omega + \omega} \, ,
	\quad
	\mu = \frac{1}{2}(\Omega - \omega) \, ,
	\quad
	\nu = - \frac{1}{2}(\Omega + \omega) \, .
	\eeq
Therefore one has a periodic and closed curve iff
	\beq \label{periodicity-const}
	\frac{\nu}{\mu} = \frac{\omega + \Omega}{\omega - \Omega} \in \mathbb{Q} \, .
	\eeq
It is clear that this condition is definitely fulfilled for $\Omega^2 + |\xi|^2 = \omega^2$, if $(\Omega, |\xi| ,\omega)$ are \emph{Pythagorean triples}, but to fulfil \eqref{periodicity-const} it is also sufficient if $\omega$ and $\Omega$ are both integers or both half-integers.
Some examples of these periodic cases are displayed in Fig. \ref{Fig-4}.
	
For this system, the auxiliary variables $(Q, P)$ evolve according to the linear system of equations
	\beq  \label{Q-P-Exp}
	\left(
		\begin{array}{ccc}
			\dot{Q} \\  
			\\
			\dot{P}
		\end{array}
	\right)
	 = 
	\left(
		\begin{array}{cc}
		2 \kappa \, (\beta_{\tiny \mbox{I}} \cos{\omega t} - \beta_{\tiny \mbox{R}} \sin{\omega t})
		& 1 - 2 \frac{ \kappa }{ \omega} \, (\beta_{\tiny \mbox{R}} \cos{\omega t} 
	+ \beta_{\tiny \mbox{I}} \sin{\omega t})  \\		
		& \\
		- \omega^2 
		- 2 \kappa \, \omega \, (\beta_{\tiny \mbox{R}} \cos{\omega t} + \beta_{\tiny \mbox{I}} \sin{\omega t}) 
		& - 2 \kappa \, (\beta_{\tiny \mbox{I}} \cos{\omega t} - \beta_{\tiny \mbox{R}} \sin{\omega t}) \\
		\end{array}
	\right)
	\left(
		\begin{array}{ccc}
			Q \\  
			\\
			P
		\end{array}
	\right) \, ,
	\eeq
where the solutions are such that they must obey the constraint $\bar{Q}(t) P(t) - Q(t) \bar{P}(t) = 2 \i$.
One may prove by direct substitution that the solution of the linear system of equations \eqref{Q-P-Exp} is given by
	\begin{align} \label{Q-solution}
	Q(t) & = a(t) Q_0 + b(t) P_0 \nonumber \\
	& = \frac{Q_0}{2} 
	\Bigg[ 
	\left( 1 - \frac{\omega}{\Omega} \right)\cos\left\{ \frac{\Omega - \omega}{2} t \right\} +
	\left( 1 + \frac{\omega}{\Omega} \right)\cos\left\{ \frac{\Omega + \omega}{2} t \right\} 
	+ \frac{4 \kappa \varrho}{\Omega}
	\Big( \cos\left\{ - \frac{\Omega + \omega}{2} t + \theta \right\} \nonumber \\
	& - \cos\left\{ \frac{\Omega - \omega}{2} t + \theta \right\} \Bigg)
	\Bigg]
	+ \frac{P_0}{2 \, \omega}\Bigg[ 
	\left( 1 + \frac{\omega}{\Omega} \right)\sin\left\{ \frac{\Omega + \omega}{2} t \right\} 
	- \left( 1 - \frac{\omega}{\Omega} \right)\sin\left\{ \frac{\Omega - \omega}{2} t \right\} \nonumber \\
	&
	+ \frac{4 \kappa \varrho}{\Omega}
	\Big( \sin\left\{ - \frac{\Omega + \omega}{2} t + \theta \right\} 
	- \sin\left\{ \frac{\Omega - \omega}{2} t + \theta \right\} \Bigg)
	\Bigg]
	\end{align}
and similarly
	\begin{align} \label{P-solution}
	P(t) & = c(t) Q_0 + d(t) P_0 \nonumber \\
	& = \frac{\omega \, Q_0}{2}\Bigg[ 
	\left( 1 - \frac{\omega}{\Omega} \right)\sin\left\{ \frac{\Omega - \omega}{2} t \right\} 
	- \left( 1 + \frac{\omega}{\Omega} \right)\sin\left\{ \frac{\Omega + \omega}{2} t \right\}
	+ \frac{4 \kappa \varrho}{\Omega}
	\Big( \sin\left\{ - \frac{\Omega + \omega}{2} t + \theta \right\} \nonumber \\
	&- \sin\left\{ \frac{\Omega - \omega}{2} t + \theta \right\} \Bigg)
	\Bigg] + \frac{P_0}{2} 
	\Bigg[ 
	\left( 1 - \frac{\omega}{\Omega} \right)\cos\left\{ \frac{\Omega - \omega}{2} t \right\} +
	\left( 1 + \frac{\omega}{\Omega} \right)\cos\left\{ \frac{\Omega + \omega}{2} t \right\} \nonumber \\
	&+ \frac{4 \kappa \varrho}{\Omega}
	\Big( \cos\left\{ \frac{\Omega - \omega}{2} t + \theta \right\} 
	- \cos\left\{ - \frac{\Omega + \omega}{2} t + \theta \right\}  \Bigg)
	\Bigg] \, ,
	\end{align}
where in the last expression $\beta = \varrho \, \e^{\i \theta}$ and the initial conditions $(Q_0, P_0)$ have been used.
Besides, one may prove that, if these solutions shall satisfy
	\beq
	\bar{Q}(t) P(t) - Q(t) \bar{P}(t) = \bar{Q}_0 P_0 - Q_0 \bar{P}_0 \, ,
	\eeq
one has to choose the initial conditions $(Q_0, P_0)$ such that our solutions satisfy the constraint $\bar{Q}(t) P(t) - Q(t) \bar{P}(t) = 2 \, \i$.

	\begin{figure}[! t]
	\centering
	\includegraphics[width = 15 cm]{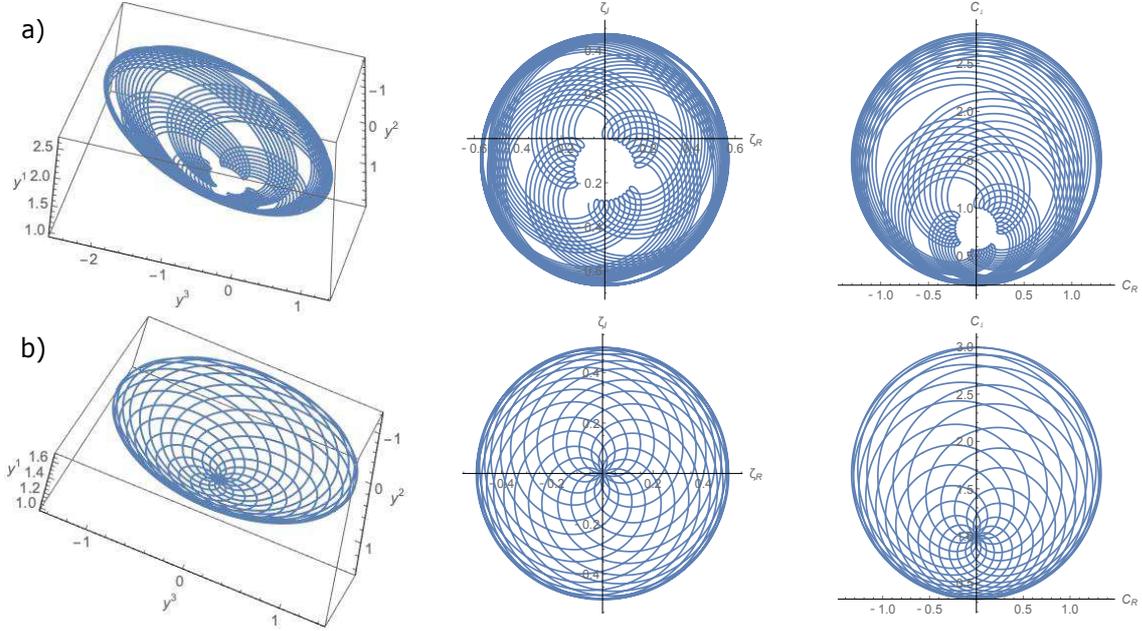}
	\caption{Time evolution on the hyperboloid $\mathbf{H}^2$ (first column), 
	in the Poincar\'e disk $\mathbb{D}^2$ (second column) and in 
	the Siegel upper half $\mathbb{HP}^2$ (third column), respectively. 
	In the last two cases the initial conditions $(Q_0, P_0) = ( 1, \i )$  are considered, but in 
	a) the parameters are $(\omega, |\xi|, \Omega) = (3/4, 1/2, \sqrt{5}/4)$, 
	while in b) the parameters are $(\omega, |\xi|, \Omega) = (1, 1/2, \sqrt{3}/4)$.}
	\label{Fig-5}
	\end{figure}

With the help of the auxiliary variables $(Q,P)$, one is able to obtain the nonlinear dynamical description of the coherent states.
To obtain the evolution on the hyperboloid $\mathbf{H}^2$, one may consider the connections in Eqs. \eqref{comp-hyp} and \eqref{Hyp-coord}. 
Some examples of this evolution are displayed in the first column in Fig.~\ref{Fig-5}.

In addition to the evolution of the coordinates $(y^1, y^2, y^3)$, its projection in the Poincar\'e disk is obtained by the transformation in Eq. \eqref{Disk-Projection}.
So, in the second column of Fig.~\ref{Fig-5}, one can find the evolution in the Poincar\'e disk.
Finally, in general the solution of the associated Riccati equation in the Siegel upper half plane is a M\"obius transformation of the form
	\beq
	\Phi(S, \Ri_0) \mapsto \Ri(t) = \frac{a(t) \Ri_0 + b(t)}{c(t) \Ri_0 + d(t)}
	\quad
	\text{with}
	\quad
	S =
	\left(
	\begin{array}{cc}
	 a(t) & b(t)  \\
	 c(t) & d(t)  
	\end{array}
	\right) \, ,
	\eeq
with $\Ri_0 = \frac{P_0}{Q_0}$ being the initial condition and  the time-dependent entries $a(t)$, $b(t)$, $c(t)$ and $d(t)$ of the symplectic matrix $S$ are given in Eqs. \eqref{Q-solution} and \eqref{P-solution}.
Some examples of the solutions of the Riccati equation are displayed in the third column of Fig~\ref{Fig-5}.

%%%%%%%%%%%%%%%%%%%%%%%%%%%%%%%%%%%%%%%%%%%%%%%%%%%%%%%%%%%%%%%%%%%%%%%%%%%%%%%%%%%%%%%%%%%%%%%%%%%%%%%%%%%%%%%%%%%%%%%%%%%%%%%%%%%%%%%%%%%%%%%%%%%%%%%%%%%%%%%%%%%%%%%%

\section{Conclusions and Perspectives}

\label{Section-6}

In this work a geometrical study of the generalized coherent states evolving under the action of Hamiltonian operators quadratic in position and momentum variables has been considered.
To perform the study of this kind of systems, the fact that in Quantum Mechanics it is possible to immerse a ``classical'' manifold into the Hilbert space has been employed, such that the evolution of the wave function is described by the evolution on this manifold.
Besides, the motion of the wave function has been restricted to lie in a certain predetermined region of the Hilbert space. 
The advantage of this immersion is that there is an interplay between quantum and classical concepts that allows to consider the procedures and structures available in the classical theory to be employed in Quantum Mechanics.

We started the study realizing that the evolution of Gaussian wave packets is encoded in the evolution of the expectation values $( \langle \hat{q} \rangle, \langle \hat{p} \rangle )$ and the time-dependent complex functions $(Q,P)$, such that these states may be expressed as in \eqref{Gaussian-WP} (or \eqref{P-Gaussian-WP}). 
So, these two aspects of the generalized coherent states reside in different manifolds. 
On the one side the expectation values $( \langle \hat{q} \rangle, \langle \hat{p} \rangle )$ live in a Euclidean linear phase space $\R^2$ with Hamiltonian evolution, according to the Ehrenfest theorem.
On the other hand, the parameters $(Q,P)$ live in the manifold $M$ defined in Eq.~\eqref{M-def} and are directly connected with the dispersions and correlation $(\sigma_q, \sigma_p, \sigma_{qp})$ as is shown in Eq. \eqref{Unc-QP}; besides, the dynamics in $M$ is Hamiltonian where the evolution equations have the same form as the classical equations of motion, but with complex variables, see Eq.~\eqref{Eq-Com-Var}.

However, the before mentioned parametrization of the generalized coherent states is not unique. 
Recall that in Quantum Optics it is possible to describe these states by the so-called Squeezing Parameters $(\tau, \varphi)$, where such parameters are the system of coordinates adapted to the hyperboloid $\mathbf{H}^2$, which is connected with $(\sigma_q, \sigma_p, \sigma_{qp})$ by Eq. \eqref{Unc-Sq}. 
The dynamical properties on $\mathbf{H}^2$ have also  been analized,  showing that there is a symplectic structure defined on it and allows to see that one has a Hamiltonian dynamics which in addition is nonlinear.  

To show the connection between the descriptions in $M$ and on $\mathbf{H}^2$ an intermediate step was necessary. This is, first one has to show the connection between $M$ and the hyperboloid $\mathbf{H}^3$ by Eq.~\eqref{comp-hyp}; thus, the connection between the hyperboloids $\mathbf{H}^3$ and $\mathbf{H}^2$ turns out to be a mere consequence of the relation between the special linear group $\text{SL}(2, \R)$ and its Lie algebra $\sl(2, \R)$, giving rise to the covering map in Eq.~\eqref{Hyp-coord}.

Finally, the last part of this work was devoted to the nonlinear Riccati dynamics. 
Taking into account the symmetry of the dynamics $X_M \in \X(M)$ under the multiplication by a global phase factor one may reduce $M$ to a lower dimensional space known as the Siegel upper half plane $\HP^2$ defined  as the space of complex numbers with strictly positive imaginary part. 
So, one not only is able to reduce the space but also it is possible to project the dynamics from $M$ onto $\HP^2$, where the projected dynamics is the nonlinear Riccati evolution, which is a Hamiltonian dynamics.
In this process of reduction one also may show a completely new parametrization of the Gaussian wave packets by means of the points in the Poincar\'e disc $\mathbb{D}^2$, which turns out to be the stereographic projection of $\mathbf{H}^2$ onto the plane, see Fig.~\ref{Fig-3}a, and whose dynamics is also Hamiltonian with Riccati-type evolution.
All the analized parametrizations and their connections are summarized in the following diagramme
	
	\begin{diagram}
	M		& 	\rTo_{\nu} 	&	\mathbf{H}^3     \\
			&			&      \dTo_{\chi}	      	 \\
	\dTo_{\pi} &			&	\mathbf{H}^2	\\ 
			&			&     	\dTo_{v}		 \\
	\HP^2	&	\lTo_{u}	& 	\mathbb{D}^2
	\end{diagram}

It is interesting to see all the geometry involved in the kinematics and the dynamics of this physical system. 
We have shown that there is a linear and Hamiltonian dynamical description of the uncertainties and correlation of a Gaussian wave packet in the space $\mathbf{H}^3$ (or equivalently in $M$); however, taking into account the symmetry associated to the action of the group $U(1)$ we ended in the lower dimensional space $\mathbf{H}^2$, here the dynamics is nonlinear and Hamiltonian.
Furthermore, note that all the different nonlinear equations in $\mathbf{H}^2$, i.e. Eqs. \eqref{Hyp-ham-eq}, \eqref{Osc-Ric-Q} and \eqref{Riccati-Disk}, are simply obtained from different adapted coordinates of  $\mathbf{H}^2$ and, because all these coordinates preserve the symplectic structure of the space, the connection between them are canonical transformations (or symplectomorphism). 
Finally, because the Weyl map is equivariant with respect to canonical transformations transformations, and as in two dimensions the symplectic structure coincides with the volume form form and is also equivariant with respect to the special linear group (this is no more true in higher dimensions), all these different adapted coordinates of $\mathbf{H}^2$ can parametrize the Gaussian wave packets.

The generalization to more degrees freedom of the reduction $M \to \HP$ has been considered in Ref.~\cite{ohsawa2015}.
In this reference it has been proven that the generalized gaussian wave packet has the parametrization 
	\beq
	\psi(\mathbf{q},t) = \frac{1}{(\pi \hbar)^{n/4}}\frac{1}{\sqrt{\det \mathbf{Q}}}
	\exp \left\{
	\frac{\i}{2 \hbar} (\mathbf{q} - \langle\hat{\mathbf{q}} \rangle)^{\tiny \mbox{T}} \frac{\mathbf{P}}{\mathbf{Q}} 
	(\mathbf{q} - \langle \hat{\mathbf{q}} \rangle) 
	+ \frac{\i}{\hbar} \langle \hat{\mathbf{p}} \rangle \cdot (\mathbf{q} - \langle\hat{\mathbf{q}} \rangle)
	+ \frac{\i}{\hbar} S(t)
	\right\} \, ,
	\eeq 
where $S(t)$ is a time-dependent phase and $(\mathbf{Q}, \mathbf{P})$ are complex $(n \times n)$-dimensional matrices in $\mathbf{M}$ defined as
	\beq
	\mathbf{M} =
	 \{ (\mathbf{Q}, \mathbf{P}) | \bar{\mathbf{Q}} \mathbf{P} - \bar{\mathbf{P}} \mathbf{Q} = 2 \, \i \, \mathbf{I} \} 	\, .
	\eeq

%In this reference it is showed that given the Schr \"odinger equation
%	\beq
%	\i \, \hbar \frac{\partial \psi}{\partial t} = - \frac{\hbar}{2m} \nabla^2 \Psi + V(\mathbf{q}) \psi
%	\eeq  
%where $\mathbf{q} \in \R^n$ and $\nabla^2$ stands for the Laplacian in $\R^n$, it has been proven in Ref.~\cite{hagedorn1980} that the Gaussian wave packet solution has the parametrization
%	\beq
%	\psi(\mathbf{q},t) = \frac{1}{(\pi \hbar)^{n/4}}\frac{1}{\sqrt{\det \mathbf{Q}}}
%	\exp \left\{
%	\frac{\i}{2 \hbar} (\mathbf{q} - \langle\hat{\mathbf{q}} \rangle)^{\tiny \mbox{T}} \frac{\mathbf{P}}{\mathbf{Q}} 
%	(\mathbf{q} - \langle \hat{\mathbf{q}} \rangle) 
%	+ \frac{\i}{\hbar} \langle \hat{\mathbf{p}} \rangle \cdot (\mathbf{q} - \langle\hat{\mathbf{q}} \rangle)
%	+ \frac{\i}{\hbar} S(t)
%	\right\} \, ,
%	\eeq 
%where $(\mathbf{Q}, \mathbf{P})$ are complex $(n \times n)$-dimensional matrices in $\mathbf{M}$ defined as
%	\beq
%	\mathbf{M} =
%	 \{ (\mathbf{Q}, \mathbf{P}) | \bar{\mathbf{Q}} \mathbf{P} - \bar{\mathbf{P}} \mathbf{Q} = 2 \, \i \, \mathbf{I} \} 	\, .
%	\eeq

%In order $\psi(\mathbf{q},t)$ to be solution of the Schor\"odinger equation, these matrices must satisfy the system of equations
%	\begin{align}
%	\dot{\mathbf{Q}} & =\frac{\mathbf{P}}{m} \, , \nonumber \\
%	\dot{\mathbf{P}} & = - \nabla^2 V \, (\mathbf{Q}) \, ,
%	\end{align}
%and the time dependent function $S(t)$ stands for the classical action 
%	\beq
%	S(t) = \int^t_{t_0} \left( \frac{\mathbf{p}^2(\tau)}{2 m } - V(\mathbf{q}(\tau)) \right) \d \tau \, .
%	\eeq

On the other hand, it has been proven in Refs.~\cite{heller1975, hagedorn1980, ohsawa2015} that it is also possible to parametrize the Gaussian wave packet as
	\beq
	\psi(\mathbf{q},t) = \left(\frac{\det \Ri_{\tiny \mbox{I}}}{(\pi \hbar)^{n}} \right)^{1/4}
	\exp \left\{
	\frac{\i}{2 \hbar} (\mathbf{q} - \langle \hat{\mathbf{q}} \rangle)^{\tiny \mbox{T}} \Ri
	(\mathbf{q} - \langle \hat{\mathbf{q}} \rangle) 
	+ \frac{\i}{\hbar}\langle \hat{\mathbf{p}} \rangle\cdot (\mathbf{q} - \langle \hat{\mathbf{q}} \rangle)
	+ \frac{\i}{\hbar} \phi(t)
	\right\}
	\eeq
with $\phi(t)$ a time dependent phase and where in this generalization the $(n \times n)$-dimensional matrix $\Ri = \Ri_{\tiny \mbox{R}} + \i \, \Ri_{\tiny \mbox{I}}$ is an element of the \emph{Siegel upper half space}~\cite{siegel1943, mcduff2017} defined as
	\beq
	\HP = \{ \Ri | \Ri^{\tiny \mbox{T}} = \Ri,  \Ri_{\tiny \mbox{I}} > 0 \} \, .
	\eeq
%which satisfies the matrix Riccati equation
%	\beq
%	\dot{\Ri} = - \frac{1}{m} \Ri ^2 - \nabla^2 V(\mathbf{q})
%	\eeq	
%and the time-dependent phase is given by
%	\beq
%	\phi(t) = S(t) - \int_{t_0}^t \tr\{ \Ri_{\tiny \mbox{I}}(\tau) \} \d \tau \, .
%	\eeq	
The connection between the symplectic spaces $\mathbf{M}$ and $\HP$ was already established by Siegel in 1943~\cite{siegel1943} and is given by the submersion
	\beq
	\pi : \mathbf{M} \to \HP : (\mathbf{Q}, \mathbf{P}) \mapsto \Ri = \frac{\mathbf{P}}{ \mathbf{Q}}\, ,
	\eeq
which generalizes the Riccati transformation to more degrees of freedom.
From a geometrical point of view it is shown in Ref.~\cite{ohsawa2015} that the submersion $\pi$ is actually a projection that allows to reduce the dynamics in $ \mathbf{M}$ onto the dynamics in $\HP$; for further details see Ref.~\cite{ohsawa2015}. 

%-----------------------------------------------------------------------------------------------------------------------------------------------------

\section*{Acknowledgements}

H. Cruz-Prado is grateful for the scholarship provided by CONACyT M\'exico, with reference number 379177.
G. M. would  like  to  thank  the  support provided by the Santander/UC3M Excellence Chair Programme 2019/2020, and in addition he is  a member of the Gruppo Nazionale di Fisica Matematica (INDAM), Italy.

\end{document}